\documentclass[a4paper,11pt]{article}
\usepackage{jheppub}
\usepackage[utf8x]{inputenc}
\usepackage[T1]{fontenc}
\usepackage{braket}
\usepackage{physics}
\usepackage{color}
\usepackage{amsfonts}
\usepackage{amssymb}
\usepackage{amsmath}
\usepackage[normalem]{ulem}
\usepackage[font={footnotesize}]{caption}
\usepackage{graphicx}
\usepackage{epstopdf}
\usepackage{enumitem}
\usepackage{subcaption}
\usepackage{comment}
\usepackage{mathtools}
\usepackage{ragged2e}
\usepackage{hyperref}
\hypersetup{
	colorlinks=true,
	linkcolor=blue,
	filecolor=cyan, 
	urlcolor=blue,
	citecolor=red,
} 
\usepackage[titletoc,toc,title]{appendix}
\numberwithin{equation}{section}
\usepackage[nameinlink]{cleveref}
\usepackage{float}
\usepackage{tikz}

\makeatletter
\newcommand{\vast}{\bBigg@{3}}
\newcommand{\Vast}{\bBigg@{5}}
\makeatother

\newcommand{\zbar}{\raisebox{0.2ex}{--}\kern-0.6em Z}

\renewcommand{\v}[1]{\ensuremath{\mathbf{#1}}} 
\renewcommand{\d}[2]{\frac{d #1}{d #2}} 
\let\baraccent=\= 
\renewcommand{\=}[1]{\stackrel{#1}{=}} 

\def\CM{{\cal M}}

\def\T{{\cal T}}

\def\CW{{\cal W}}

\def\tbx{{\textbf{x}}}
\def\tby{{\textbf{y}}}

\def\pab{\partial\textbf{a}}
\def\ab{\textbf{a}}

\def\d{\textrm{d}}

\def\tr{\mathrm{Tr}}

\def\D{\mathrm{d}}

\title{Holography and Kinematic Space for Gravitational Sub-regions in AdS}

\author[a]{Debarshi Basu} \author[a]{and Qiang Wen}

\affiliation[a]{School of Physics and Shing-Tung Yau Center, Southeast University, Nanjing 210096, China}

\emailAdd{debarshi.128@gmail.com} \emailAdd{wenqiang@seu.edu.cn}

\abstract{It is well-known in integral geometry that a maximally symmetric Riemannian manifold, such as a static slice of vacuum AdS spacetime, can be perfectly covered by the geodesics in the Kinematic space, which we call the partial-entanglement-entropy (PEE) threads. In this context, the area of a codimension-one surface in the manifold can be computed by counting its intersections with the PEE threads, which is the celebrated Crofton formula. In this paper, we analyze the Kinematic space for a generic subregion in vacuum AdS space, and propose that the PEE threads emanate from a co-dimension one surface can perfectly cover a subregion in the manifold. Furthermore, we build holographic tensor network models on the network of the PEE threads confined in a subregion, thereby providing a concrete framework that realizes the surface-state correspondence and the generalized entanglement wedges for gravitational subregions.
}

\begin{document} 
	\maketitle
	\flushbottom
	
\section{Introduction}
\label{sec:introduction}

The AdS/CFT correspondence provides a non-perturbative framework in which a gravitational theory in an asymptotically AdS spacetime is encoded in a quantum field theory living on its non-gravitating boundary \cite{Maldacena:1997re,Gubser:1998bc,Witten:1998qj}. Although the correspondence is conventionally formulated as a duality between two complete theories, much of its practical and conceptual power comes from a more local question: how is the information associated with a specified region distributed and reconstructed across the duality? Answering this question for boundary subregions has led to some of the sharpest connections between quantum information and spacetime geometry. A more ambitious problem is to formulate an analogous description directly for finite \emph{gravitational subregions}, whose boundaries lie partly or entirely in the bulk and need not support an autonomous non-gravitating quantum field theory. Understanding whether such regions admit an intrinsic information-theoretic description is an important step toward a genuinely local formulation of holography.

The central bridge between geometry and quantum information in AdS/CFT is the Ryu--Takayanagi (RT) formula \cite{Ryu:2006bv,Ryu:2006ef,Lewkowycz:2013nqa}. For a static boundary region $A$, it relates the von Neumann entropy of the reduced density matrix $\rho_A$ to the area of the bulk minimal surface $\mathcal{E}_A$ homologous to $A$,
\begin{align}
	S_A=-\tr\!\left(\rho_A\log\rho_A\right)
	=\frac{\operatorname{Area}(\mathcal{E}_A)}{4G}\,.
\end{align}
Its covariant extension replaces $\mathcal{E}_A$ by an extremal surface \cite{Hubeny:2007xt,Dong:2016hjy}, while quantum corrections supplement the area term by the entropy of bulk quantum fields and ultimately lead to the quantum-extremal-surface prescription \cite{Faulkner:2013ana,Engelhardt:2014gca}. These developments suggest that spatial geometry is not merely an auxiliary background for the boundary theory: geometric areas themselves encode patterns of quantum entanglement.

This viewpoint has motivated a broad program aimed at recovering the organization of the bulk from entanglement data \cite{VanRaamsdonk:2009ar,VanRaamsdonk:2010pw}; see \cite{Rangamani:2016dms,Chen:2021lnq} for reviews. A particularly useful lesson of this program is that the RT entropy represents only a coarse summary of a much richer geometric and entanglement structure. To reconstruct local bulk information, one must refine the entropy associated with a complete boundary region into quantities that resolve how different degrees of freedom are correlated and how those correlations are distributed through the bulk.

Several complementary frameworks implement this refinement. Differential entropy reconstructs the length of a closed bulk curve in AdS$_3$ by combining the entanglement entropies of a continuous family of boundary intervals \cite{Balasubramanian:2013lsa,Headrick:2014eia,Czech:2014wka,Czech:2015qta}. Integral geometry and kinematic space provide a systematic interpretation of this construction: on a static slice of vacuum AdS, the space of boundary-anchored geodesics carries a canonical measure, and the area of a bulk hypersurface can be evaluated by counting its intersections with these geodesics through the Crofton formula \cite{santalo2004integral,Czech:2015qta,Czech:2016xec}. Entanglement also constrains bulk dynamics. In particular, the first law of entanglement entropy for ball-shaped boundary regions is equivalent, under suitable assumptions, to the linearized gravitational equations in the bulk \cite{Lashkari:2013koa,Faulkner:2013ica}. Together, these results indicate that a sufficiently organized collection of entanglement data can encode both the geometry and the dynamics of a gravitational spacetime.

Tensor-network models make this intuition more concrete. In these models, the connectivity of the network represents an emergent spatial geometry, while the entanglement carried by its internal legs reproduces RT-like entropy relations, quantum error correction, and the reconstruction of bulk operators \cite{Swingle:2009bg,evenbly2011tensor,Pastawski:2015qua,Hayden:2016cfa,Bhattacharyya:2016hbx,Bhattacharyya:2017aly}. These constructions have clarified many structural aspects of holography, but most conventional tensor networks begin with a chosen discretization of the bulk. Their network geometry is therefore an input of the model rather than a structure derived directly from a continuum entanglement measure.

The bit-thread formulation offers a different continuum description of holographic entanglement \cite{Freedman:2016zud,Headrick:2020gyq,Headrick:2022nbe}. The entropy $S_A$ is expressed as the maximal flux through $A$ of a divergenceless vector field whose norm is bounded by $1/(4G)$. Integral curves of this vector field may be viewed as threads connecting $A$ with its complement $\bar A$, while the RT surface is the bottleneck saturated by a maximizing flow. This prescription reveals how the entropy can be distributed throughout the entanglement wedge and makes several entropy inequalities geometrically transparent. Nevertheless, a given boundary region generally admits infinitely many maximizing flows. Bit threads therefore provide a family of valid geometric representations of $S_A$, rather than a unique thread configuration determined by the microscopic entanglement structure.

A more intrinsic thread configuration arises from the partial entanglement entropy (PEE), an additive measure designed to quantify the contribution of a subset of degrees of freedom to the entanglement associated with a larger region \cite{Wen:2018whg,Wen:2019iyq,Wen:2020ech,Han:2019scu}. The PEE and related entanglement-contour constructions have been investigated in a variety of static, covariant, mixed-state, black-hole, and island configurations \cite{Kudler-Flam:2019oru,Wen:2018mev,Han:2021ycp,Ageev:2021ipd,Wen:2021qgx,Camargo:2022mme,Basu:2023wmv,Lin:2023ajt}. In the vacuum of a holographic CFT, the two-point PEE $\mathcal{I}(\tbx,\tby)$ can be geometrized by a bulk geodesic connecting the boundary points $\tbx$ and $\tby$ \cite{Lin:2023rxc}. We refer to this geodesic, together with its density fixed by $\mathcal{I}(\tbx,\tby)$, as a \emph{PEE thread}. In contrast with a maximizing bit-thread flow, the full PEE-thread configuration is determined by the state and does not have to be chosen separately for each entangling region. PEE threads may intersect, and their total crossing through a surface counts intersections rather than an oriented net flux.

The collection of all PEE threads on a static slice of vacuum AdS forms a continuous PEE network that uniformly tessellates the bulk geometry \cite{Lin:2024fze}. The density of this network is precisely such that the area of any codimension-one surface on the slice can be obtained by counting its intersections with the PEE threads. This makes the PEE network a natural continuum foundation for holographic tensor-network models. In particular, tensor networks constructed directly on the PEE network reproduce the RT formula because the number of network bonds intersected by an RT surface is proportional to its area \cite{Wen:2025gui}. The relation between two-point PEE, bulk geodesics, kinematic space, and the Crofton measure therefore provides a common language connecting entanglement contours, integral geometry, thread configurations, and tensor networks.

Previous investigations of this construction have primarily treated the PEE network associated with an entire static slice of Poincar\'e AdS \cite{Lin:2023rxc,Lin:2024fze,Wen:2025gui}. The purpose of the present work is to localize this framework. Given a finite region $\ab$ inside a static slice of vacuum AdS, we ask whether one can define a \emph{subregion kinematic space} $\mathbb{K}_{\ab}$ whose geodesics uniformly cover $\ab$, reconstruct the geometry inside $\ab$, and support a tensor-network model intrinsic to that region. We then address the inverse problem: starting with a codimension-one bulk surface $\Sigma$, what is the maximal region that can be reconstructed from the PEE threads emanating from $\Sigma$?

For a connected subregion $\ab$, the relevant elements of $\mathbb{K}_{\ab}$ are geodesic chords with endpoints on $\partial\ab$. Equivalently, they are the connected portions inside $\ab$ of the complete geodesics belonging to the global kinematic space. This distinction becomes important when $\partial\ab$ is non-convex, because a single complete geodesic can enter and leave $\ab$ several times and consequently define more than one chord in $\mathbb{K}_{\ab}$. It is therefore more natural to parameterize the subregion kinematic space by the endpoints of the chords on $\partial\ab$, rather than by the asymptotic endpoints of their global extensions.

In AdS$_3$, if $s_1$ and $s_2$ label two points on $\partial\ab$ and $\ell(s_1,s_2)$ denotes the length of the geodesic chord joining them, we show that the kinematic measure is
\[
	\D\Gamma_{\ab}
	=\frac{1}{2}\frac{\partial^2\ell(s_1,s_2)}
	{\partial s_1\partial s_2}\,\D s_1\wedge\D s_2\,.
\]
We subsequently derive its higher-dimensional generalization from an intrinsic measure on the space of intersections between geodesics and a hypersurface. If $x^i$ are coordinates on a codimension-one surface $\Sigma$ and $p_i$ are the tangential components of the canonical geodesic momentum, the measure
\[
	\D\Gamma'_{\Sigma}
	=\D p_1\wedge\D x^1\wedge\cdots
	\wedge\D p_{d-1}\wedge\D x^{d-1}
\]
defines a distribution of intersections that is uniform over $\Sigma$ and isotropic in direction. For manifolds admitting a global kinematic space, we demonstrate that this intrinsic measure agrees with the measure induced by the global Crofton distribution. Applying it to $\partial\ab$ produces the kinematic measure on $\mathbb{K}_{\ab}$. Consequently, the area of any codimension-one surface contained in $\ab$ can be reconstructed by counting its intersections with the geodesic chords in $\mathbb{K}_{\ab}$.

This construction also gives a precise geometric meaning to reconstruction from a bulk surface. Let $C_{\Sigma}$ denote the family of PEE threads shot from $\Sigma$ with the intrinsic intersection measure. We define the reconstruction region $W_{\Sigma}$ as the set of points for which the complete local Crofton distribution is contained in $C_{\Sigma}$. Equivalently, every geodesic passing through a point of $W_{\Sigma}$ must belong to the family emitted from $\Sigma$. The area of every hypersurface inside $W_{\Sigma}$ can then be reconstructed solely from its intersections with $C_{\Sigma}$. This is initially a statement of integral geometry, rather than a claim of exact operator reconstruction in a microscopic theory of quantum gravity. It nevertheless supplies the geometric structure needed to construct explicit models of holography for gravitational subregions.

The boundary of $W_{\Sigma}$ is naturally controlled by totally geodesic submanifolds. Such surfaces act as barriers because a geodesic intersecting them either crosses only once or lies entirely within them. We analyze reconstruction from closed convex surfaces, open convex surfaces, and generic non-convex surfaces. A closed convex surface reconstructs precisely the region it encloses. An open convex surface can reconstruct the region bounded by itself and an appropriate totally geodesic completion. A non-convex surface may reconstruct regions on both of its sides, with different totally geodesic segments closing the corresponding components. For a spherical region on the asymptotic boundary, the completing surface is its RT hemisphere, and $W_{\Sigma}$ coincides with the usual entanglement wedge. Through the AdS-Rindler construction, the same reasoning shows that the exterior of a BTZ black brane, and likewise the exterior of the BTZ black hole, can be reconstructed from the asymptotic boundary because the horizon is totally geodesic \cite{Casini:2011kv}.

The subregion PEE network also allows us to construct tensor-network models whose effective boundary is a surface inside the gravitational spacetime. This provides a concrete realization of several proposals collectively known as the surface/state correspondence \cite{Miyaji:2015yva,Miyaji:2015fia,Chen:2019mis,Bao:2023til}. In the factorized PEE tensor network, a topologically trivial closed surface carries a pure state formed from the EPR pairs represented by PEE threads with both endpoints on the surface. An open surface carries a mixed state obtained by tracing out the degrees of freedom on its complement. Its entropy counts the PEE threads connecting the surface to that complement and is therefore determined by the area of the homologous geodesic. For a topologically non-trivial surface enclosing a BTZ horizon, tracing over the horizon legs similarly produces a mixed state whose entropy is fixed by the horizon area.

The model further clarifies an important subtlety in the proposed relation between homologous bulk surfaces. Surfaces of different areas support different numbers of tensor-network legs and therefore do not literally define identical Hilbert spaces. The deformation between them acts unitarily on the degrees of freedom that can be matched by extending the same PEE threads, while additional internal EPR pairs disappear as the surface is contracted. These pairs do not contribute to the entropy across the relevant bipartition. The PEE network thus realizes the entropy-preserving content of the surface/state correspondence while making explicit the assumptions required to compare states assigned to different bulk surfaces.

Finally, we relate the reconstruction region determined by the subregion kinematic space to the generalized entanglement wedge proposed for gravitating regions \cite{Bousso:2022hlz,Bousso:2023sya,Kaya:2025vof,Arayath:2026rll,Geng:2020fxl,Geng:2023qwm}. In contrast with the ordinary entanglement wedge, whose input is a region on a non-gravitating asymptotic boundary, the generalized construction assigns a larger reconstructable region to a region that is itself gravitational. This idea is closely connected with the appearance of islands in the entanglement wedge of Hawking radiation \cite{Penington:2019npb,Almheiri:2019psf,Almheiri:2020cfm}. In the classical limit considered here, where bulk entanglement corrections are neglected, we show in representative configurations that the generalized wedge selected by minimizing the boundary area agrees with the region reconstructed by the PEE threads emitted from the appropriate bulk surface. The subregion PEE tensor network thereby furnishes a simple geometric model of how information stored on a gravitational region may encode a larger portion of the spacetime.

Our results should be interpreted within their present scope. The construction is developed primarily on static slices of vacuum AdS, where a canonical global Crofton measure exists. Moreover, the factorized PEE tensor network captures only pairwise EPR-like correlations. It does not, for example, reproduce every connected entanglement wedge associated with multiple disjoint boundary intervals. Incorporating genuine multipartite correlations, possibly through random tensors defined on the PEE network, is expected to be necessary for a more complete model \cite{Hayden:2016cfa,Wen:2025gui}. Extensions to geometries containing matter, generic black holes, quantum corrections, and non-AdS spacetimes will also require additional classes of PEE-thread endpoints and a corresponding generalization of the kinematic measure.

The remainder of the paper is organized as follows. Section~\ref{sec2} reviews the global kinematic space of vacuum AdS, derives the subregion kinematic measure via intrinsic intersections on $\partial\textbf{a}$ (Theorems~1--2), and analyzes reconstruction from closed convex, open convex (including the entanglement wedge and BTZ exterior), and non-convex surfaces. Section~\ref{sec3} builds factorized PEE tensor networks realizing the surface/state correspondence for closed, open, and topologically non-trivial surfaces, with the entropy of an open surface given by the area of the homologous geodesic and homologous surfaces related by unitary evolution. Section~\ref{sec4} relates the PEE reconstruction regions to generalized entanglement wedges for gravitational subregions. We conclude in Section~\ref{sec5} with limitations and future extensions to non-AdS geometries, matter fields, and quantum corrections.
\section{Kinematic space for subregions and subregion reconstruction in AdS}
\label{sec2}
\subsection{The Kinematic space for AdS space }
First, let us briefly introduce the basic ideas and concepts of integral geometry \cite{santalo2004integral}. Given a Riemannian manifold, the kinematic space is the space of all geodesics on the manifold, equipped with a measure that is invariant under the symmetries of the manifold. Such spaces exist only for highly symmetric manifolds, such as Euclidean, spherical, and hyperbolic geometries, where symmetry transformations can map any geodesic to any other. Hence, the kinematic space provides a unique and uniform covering of the manifold by its geodesics. Under such a covering, the area of any codimension-one surface can be computed by counting the number of its intersections with the geodesics in the kinematic space; this is the celebrated Crofton formula. Nevertheless, in a generic Riemannian manifold, it is not clear how to define such a canonical, invariant way to count intersections with geodesics, which makes the Crofton formula inapplicable.

More explicitly, consider a $d$-dimensional Riemannian manifold $\mathcal{M}$ for which the kinematic space $\mathbb{K}_{\mathcal{M}}$ can be constructed, and let $\Sigma$ be an arbitrary $(d-1)$-dimensional hypersurface in $\mathcal{M}$. The Crofton formula states that the area of $\Sigma$ can be evaluated by counting the number of intersections with the geodesics in $\mathbb{K}_{\mathcal{M}}$,
\begin{equation}\label{Crofton}
	\text{Area}\left(\Sigma\right)=\frac{1}{2}\frac{d-1}{\Omega_{d-2}} \int_{\mathbb{K}_{\mathcal{M}}}\#\left(\Gamma\cap \Sigma\right)d\Gamma.
\end{equation}
Here $\Gamma$ denotes an arbitrary geodesic in $\mathbb{K}_{\mathcal{M}}$, $\d\Gamma$ is the kinematic measure, $\Omega_{d-2}=\frac{2\pi^{(d-1)/2}}{\Gamma((d-1)/2)}$ stands for the area of a unit $(d-2)$-sphere, and $\#(\Gamma\cap\Sigma)$ counts how many times the geodesic $\Gamma$ pierces the surface $\Sigma$. Equivalently, the Crofton formula can be interpreted as follows: the intersection density between any bulk codimension-one surface $\Sigma$ and the geodesics in $\mathbb{K}_{\mathcal{M}}$ is constant on $\Sigma$.

The Crofton formula is now a particularly useful tool to understand AdS/CFT \cite{Czech:2015qta}, since the time slice $\mathcal{M}$ of vacuum AdS$_{d+1}$ is a hyperbolic space where the kinematic space can be constructed. In this case, all geodesics anchor on the boundary; hence, the kinematic space $\mathbb{K}_{\mathcal{M}}$ can be parameterized by the endpoint coordinates $\left\{\v{x},\v{y}\right\}$ of all the geodesics. Furthermore, the kinematic measure $d\Gamma$ is given by \cite{Czech:2015qta,Czech:2016xec}:
\begin{equation}\label{Kinematicmeasure}
	\begin{aligned}
		d\Gamma&=\det\left(\frac{\partial^2 \ell\left(\v{x},\v{y}\right)}{\partial \v{x}\partial \v{y}}\right)d\v{x}\wedge d \v{y}=\frac{2^{d-1}}{\left|\v{x}-\v{y}\right|^{2d-2}}d\v{x}\wedge d \v{y},
	\end{aligned}
\end{equation}
where $\ell\left(\v{x},\v{y}\right)$ is the length of the geodesic anchored at $\v{x}$ and $\v{y}$. 

On the other hand, it is worth mentioning that an additive entanglement structure $\mathcal{I}(\v{x},\v{y})$ called the partial entanglement entropy (PEE) \cite{Wen:2018whg,Wen:2019iyq,Wen:2020ech,Han:2019scu} \footnote{See also the studies on the extensive mutual information \cite{Casini:2008wt} in vacuum CFTs, which share the same formula as the PEE, but do not apply to holographic CFTs.} was proved to exist for the vacuum state of CFTs, and the formula for the two-point PEE is given by
\begin{equation}\label{twopointpee}
	\mathcal{I}\left(\v{x},\v{y}\right)=\frac{c}{6}\frac{2^{\left(d-1\right)}\left(d-1\right)}{\Omega_{d-2}\left|\v{x}-\v{y}\right|^{2\left(d-1\right)}}.
\end{equation}
Comparing \eqref{Kinematicmeasure} and \eqref{twopointpee}, one immediately sees that the two-point PEE is proportional to the kinematic measure: both share the same factor $|\v{x}-\v{y}|^{-2(d-1)}$. This observation suggests that the kinematic space for AdS can be reconstructed from the PEE structure through a two-step procedure:
\begin{itemize}
	\item promote each two-point PEE $\mathcal{I}(\v{x},\v{y})$ to a bulk geodesic anchored at $\v{x}$ and $\v{y}$;
	\item impose that the total number of PEE threads connecting any two non-overlapping regions $A$ and $B$ is given by the integrated PEE, $\mathcal{I}(A,B)=\int_A\d\v{x}\int_B\d\v{y}\,\mathcal{I}(\v{x},\v{y})$.
\end{itemize}
For further details we refer the reader to \cite{Lin:2023rxc,Lin:2024fze}. In this geometric picture, each geodesic in $\mathbb{K}_{\mathcal{M}}$ acquires a direct physical meaning as a PEE thread, which geometrizes the two-point entanglement structure of the boundary CFT.

Unlike \cite{Czech:2015qta}, which focuses on computing volumes in the kinematic space to reconstruct the lengths of curves in the AdS bulk, here we follow \cite{Lin:2023rxc,Lin:2024fze} to construct explicit configurations of PEE threads that cover the original AdS space. More explicitly, we use a vector field $V_{\tbx}$ to describe the PEE threads emanating from any boundary site $\v{x}$, with its norm capturing the density of the threads and its integral curves coinciding with the geodesics emanating from $\v{x}$. The explicit formula for $V_{\tbx}(Q)$ has been derived in \cite{Lin:2023rxc} for Poincar\'e AdS in general dimensions; for example, in AdS$_3$ we have:
\begin{align}
	V_{x_0}^\mu(\bar x,\bar z)=\frac{1}{4G_N}\frac{2\bar{z}^2(\bar{x}-x_0)}{\left[(\bar{x}-x_0)^2+\bar{z}^2\right]^2}\left(\bar{z},\frac{\bar{z}^2-\left(\bar{x}-x_0\right)^2}{2\left(\bar{x}-x_0\right)}\right)\label{PEE-thread-flow},
\end{align}
where $\bar{x}$ and $\bar{z}$ represent the coordinates of the bulk point $Q$. 

Furthermore, the superposition of all such vector fields forms a continuous network of PEE threads, which we call the PEE network. It is important to stress that the PEE network is an exact tessellation of the AdS space, making it a perfect structure for defining tensor network models of quantum gravity (see \cite{Wen:2025gui} for examples). Moreover, since counting the ``number''\footnote{In the PEE network, the number of geodesics is countless, but one can compute the ``flux'' associated with these vector fields. Additionally, one can compute the total crossing of the threads through a given bulk surface $\Sigma$. It is important to note that the PEE threads are unoriented curves; each time a thread intersects $\Sigma$, it contributes positively to the total crossing count. Therefore, when computing this quantity, which differs from the conventional net flux in that contributions from opposite directions add rather than cancel, the threads should be weighted by the number of times they intersect $\Sigma$.} of intersections between a bulk surface $\Sigma$ and the PEE network is equivalent to computing the area of $\Sigma$, the RT formula follows directly from a computation of the entanglement entropy for boundary regions in the holographic tensor network models \cite{Wen:2025gui}.

\subsection{Kinematic space for subregions}
\begin{figure}[ht]
	\centering
	\includegraphics[width=0.5\linewidth]{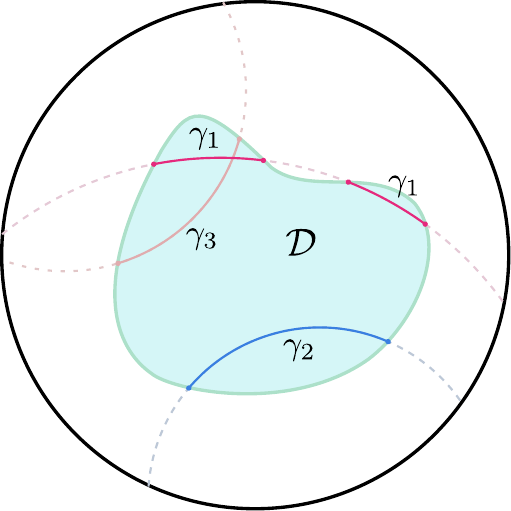}
	\caption{The disk represents a time slice of vacuum AdS$_{3}$, and the blue region represents an arbitrary subregion $\textbf{a}$. The dashed lines are geodesics in $\mathbb{K}$, and the solid lines $\gamma_i$ represent the portions of them that lie within $\textbf{a}$. These geodesic chords $\gamma_i$ are precisely the elements of $\mathbb{K}_{\textbf{a}}$ that uniformly cover $\textbf{a}$.}
	\label{fig:elementsinKa}
\end{figure}

Now we generalize the kinematic space $\mathbb{K}$ for the entire vacuum AdS space to the kinematic space $\mathbb{K}_{\textbf{a}}$ for any connected subregion $\textbf{a}$ in AdS. More explicitly, we consider any subregion $\textbf{a}$ in AdS surrounded by a closed curve (or surface) $\partial \textbf{a}$ and identify the set of geodesics, which we denote by $\mathbb{K}_{\textbf{a}}$, that uniformly cover $\textbf{a}$. The uniform covering also means that the area of any surface in $\textbf{a}$ can be reconstructed by counting the number of intersections it makes with the geodesics in $\mathbb{K}_{\textbf{a}}$. Note that the geodesics in $\mathbb{K}$ uniformly cover the whole vacuum AdS space, and their portions lying inside any $\textbf{a}$ should also uniformly cover $\textbf{a}$. In other words, the kinematic space $\mathbb{K}_{\textbf{a}}$ exists for any $\textbf{a}$, and it is simply the set of portions of the geodesics in $\mathbb{K}$ that lie within $\textbf{a}$ (see Fig.~\ref{fig:elementsinKa} for a simple illustration).

It is obvious that we have a map sending elements of $\mathbb{K}_{\textbf{a}}$ to the geodesics in $\mathbb{K}$ that intersect $\partial \textbf{a}$. However, this map is not one-to-one, since some geodesics in $\mathbb{K}$ do not enter $\textbf{a}$ at all. Moreover, when the boundary $\partial \textbf{a}$ is not convex\footnote{A surface is convex if all geodesics connecting any two points on the surface lie entirely within the region it encloses.}, a geodesic in $\mathbb{K}$ may intersect $\partial \textbf{a}$ more than twice, thus corresponding to more than one element in $\mathbb{K}_{\textbf{a}}$. For example, the geodesic chords $\gamma_1$ and $\gamma_2$ in Fig.~\ref{fig:noncovex} correspond to the same geodesic in $\mathbb{K}$. These facts make it inconvenient to parameterize $\mathbb{K}_{\textbf{a}}$ using the parameters of $\mathbb{K}$\footnote{See \cite{Czech:2015qta} for examples of parameterizing the geodesics in $\mathbb{K}$ that intersect the boundary $\partial \textbf{a}$ of certain subregions in AdS$_3$ using the parameters of $\mathbb{K}$.}. It is therefore convenient to parameterize the geodesics in $\mathbb{K}_{\textbf{a}}$ by the coordinates of their endpoints on $\partial \textbf{a}$, which is our main task in this subsection.  

Let us consider the simpler case of AdS$_3$, where $\partial \textbf{a}$ is a closed curve parameterized by a coordinate $s$. We propose that the kinematic measure for $\mathbb{K}_{\textbf{a}}$ is given by
\begin{equation}\label{eq:Gamma_a}
	\D \Gamma =  \frac{1}{2}\frac{\partial^2 \ell (s_1,s_2)}{\partial s_1 \partial s_2}  \D  s_1  \D s_2,
\end{equation}
which closely resembles the measure \eqref{Kinematicmeasure} for $\mathbb{K}$, with $\ell (s_1,s_2)$ being the length of the geodesic anchored at $s_1$ and $s_2$ on $\partial \textbf{a}$. Note that $\mathbb{K}_{\textbf{a}}$ exists and the Crofton formula holds for any curve in $\textbf{a}$. Consider any subinterval $A$ of $\partial \textbf{a}$ with $A\cup \bar{A}=\partial \textbf{a}$; the length of the geodesic $\gamma_A$ homologous to $A$ counts the number of threads connecting $A$ and $\bar{A}$, i.e.
\begin{align}
	\textit{length}(\gamma_A)=& \mathcal{N}(A,\bar{A}),
\end{align}
where $\mathcal{N}(A,\bar{A})$ denotes the number of geodesics connecting $A$ and $\bar{A}$. The proof is simple. Since $\gamma_A$ is itself a geodesic, the only geodesics that intersect $\gamma_A$ are those connecting $A$ and $\bar A$, and each such geodesic intersects $\gamma_A$ exactly once. In other words, the number $\mathcal{N}(A,\bar{A})$ equals the number of intersections between $\gamma_A$ and the geodesics in $\mathbb{K}_{\textbf{a}}$, which, by the Crofton formula, reproduces the length of $\gamma_A$. 

Now consider two infinitesimal intervals on $\partial \textbf{a}$,
\begin{align}
	A_1: s_1\to s_1+d s_1,\qquad A_2: s_2\to s_2+ds_2,
\end{align}
where $ds_1$ and $ds_2$ are infinitesimal parameters. Let $B_1,~B_2$ be the two complementary intervals such that $\partial \textbf{a}=A_1\cup A_2\cup B_1\cup B_2$. Denote the length of a geodesic by the coordinates of its boundary points; for example, $\textit{length}(\gamma_{A_1B_1})=\ell (s_1,s_2)$. Using the additivity property\footnote{For any non-overlapping regions $A, B$ and $C$, additivity states that $\mathcal{N}(A,BC)=\mathcal{N}(A,B)+\mathcal{N}(A,C)$.}, we have
\begin{align}
	\frac{\partial^2 \ell (s_1,s_2)}{\partial s_1 \partial s_2}= & \frac{\ell(s_1+ds_1,s_2+ds_2)-\ell(s_1+ds_1,s_2)}{ds_1ds_2}-\frac{\ell(s_1,s_2+ds_2)-\ell(s_1,s_2)}{ds_1ds_2} \cr
	= & \frac{\mathcal{N}(A_2B_1, A_1 B_2)-\mathcal{N}(B_1,A_1A_2B_2)}{ds_1ds_2}-\frac{\mathcal{N}(A_1A_2B_1, B_2)-\mathcal{N}(A_1B_1,A_2B_2)}{ds_1ds_2}
	\cr
	= & 2\frac{\mathcal{N}(A_1,A_2)}{ds_1 ds_2}.
\end{align}
Substituting this result into \eqref{eq:Gamma_a} yields
\begin{align}
	\D \Gamma = \mathcal{N}(A_1,A_2),
\end{align}
which exactly captures the number of geodesics connecting the two infinitesimal subintervals $ds_1$ and $ds_2$. This shows that \eqref{eq:Gamma_a} is precisely the kinematic measure for $\mathbb{K}_{\textbf{a}}$.

\begin{figure}[ht]
	\centering
	\includegraphics[width=0.5\linewidth]{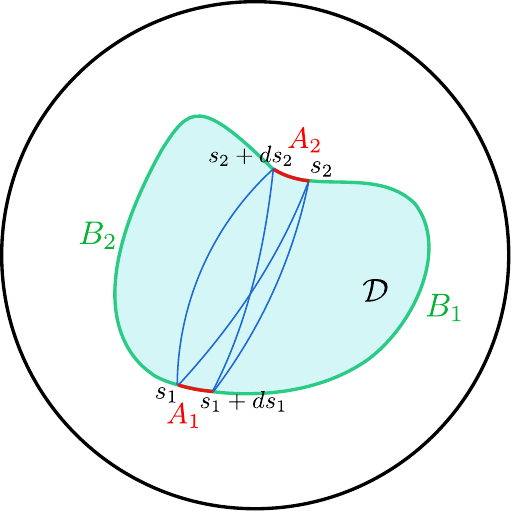}
	\caption{Illustration for the proof of the kinematic measure.}
	\label{fig:provekm}
\end{figure}

Generalizing the above discussion to higher dimensons could be possible, since the RT surfaces for spherical regions are totally geodesic submanifolds that can play the role of $\gamma_A$ in AdS$_3$. One may use these totally geodesic submanifolds to construct tubes connecting infinitesimal area elements on $\partial \textbf{a}$. Nevertheless, the generalization would be complicated, and in the following we will introduce a more general method to construct $\mathbb{K}_{\textbf{a}}$ in higher dimensions.

Let $\mathcal{M}$ be a $d$-dimensional Riemannian manifold and let $\Sigma\subset\mathcal{M}$ be any smooth codimension-one hypersurface. In a local coordinate chart adapted to $\Sigma$, we take $x^{d}=0$ as the equation defining $\Sigma$, with $(x^1,\dots,x^{d-1})$ providing coordinates along $\Sigma$. Following the discussion in Chapter 19 of \cite{santalo2004integral}, one can then introduce a measure for the space of all local intersections between geodesics in $\mathcal{M}$ and the surface $\Sigma$,
\begin{align}\label{eq:inv_dens}
	\D \Gamma'_{\Sigma} = \D p_1 \wedge \D x^1 \wedge \cdots \wedge \D p_{d-1}\wedge \D x^{d-1},
\end{align}
Here the canonical momenta $p_\mu$ are obtained from the geodesic Lagrangian
\begin{align}
	\mathcal{L} = \sqrt{g_{\mu\nu}\dot{x}^\mu \dot{x}^\nu},\qquad \dot{x}^\mu = \D x^\mu(\tau) / \D\tau, \qquad p_\mu = \frac{\partial \mathcal{L}}{\partial \dot{x}^\mu},
\end{align}	
with $x^{\mu}(\tau)$ denoting the geodesic curve parametrized by an affine parameter $\tau$. Here the coordinates $\{x^i\}$ (with $i=1,\dots,d-1$) characterize where the geodesic intersects $\Sigma$, and the momenta $\{p_i\}$ characterize the intersecting direction. In other words, an intersection is specified by both the position and the direction of the geodesic at that point.

One can determine a set of geodesics $\mathcal{C}_{\Sigma}$ intersecting $\Sigma$ by shooting geodesics from $\Sigma$ along all possible intersections with the measure \eqref{eq:inv_dens}. Note that the space of intersections is not equivalent to $\mathcal{C}_{\Sigma}$ because a single geodesic may intersect $\Sigma$ multiple times, thus corresponding to several distinct intersections. However, in principle the correspondence between geodesics and intersections can be worked out, so the measure $d\Gamma_{\Sigma}$ on $\mathcal{C}_{\Sigma}$ is determined by $d\Gamma'_{\Sigma}$, and vice versa.

We claim that the measure \eqref{eq:inv_dens} actually describes the unique distribution of intersections that is uniform over $\Sigma$ and isotropic over all directions. It can be defined for any smooth surface $\Sigma$ in a generic manifold $\mathcal{M}$. It was proved in \cite{santalo2004integral} that the measure \eqref{eq:inv_dens} is invariant under changes of coordinates and under translations along $\tau$. Furthermore, it was proved that in a neighborhood of any point $P$ on $\Sigma$ we always have\footnote{Locally, one can always choose normal coordinates such that the metric is diagonal, and $dp_i=\sqrt{g_{ii}}\sin \alpha_i d\alpha_i$. Then, following the steps in Chapter 19 of \cite{santalo2004integral}, we obtain
	\begin{align*}
		\D p_1 \wedge \D x^1 \wedge \cdots \wedge \D p_{d-1}\wedge \D x^{d-1} & =\sqrt{g_{ii}\cdots g_{d-1,d-1}}\sin\alpha_1\cdots\sin\alpha_{d-1} dx^1\wedge\cdots \wedge dx^{d-1}\wedge d\alpha_1\cdots \wedge d\alpha_{d-1}
		\cr
		& = |\cos \alpha_d|\,d\sigma\wedge d\Omega_{d-1}.
\end{align*} } 
\begin{equation}\label{eq:gammapsigma}
	\D \Gamma'_{\Sigma} = |\cos \alpha _d| \, \D \sigma \wedge \D \Omega_{d-1},
\end{equation}
where $\alpha_i$ is the angle between the geodesic and the $x_i$ direction, $d\sigma$ is the area element of $\Sigma$ at $P$, $d\Omega_{d-1}$ is the area element on the $(d-1)$-sphere, and $\alpha_d$ is the angle between the geodesic and the normal direction $x_d$. Here, $\D \sigma \wedge \D \Omega_{d-1}$  is independent of the position of $P$ and the direction of the intersections. The isotropic distribution is represented by the factor $\D \Omega_{d-1}$, the uniform distribution in space is represented by the factor $d\sigma$. The $|\cos \alpha _d|$ factor extracts the normal component of the intersections that contributes to the number (i.e., the flux of geodesics passing through $d\sigma$) of intersections\footnote{One can use vector fields to describe the distribution of geodesics, and the number of intersections is equivalent to computing the flux (or total crossing) of the vector fields through the surface; hence, only the normal component of the vector field contributes.}.  After integrating over half of the unit sphere\footnote{Here we consider only unoriented geodesics.}, we obtain a universal density $d\sigma \frac{\Omega_{d-2}}{d-1}$ at any $P$. Integrating further over $\Sigma$ gives
\begin{equation}
	\begin{aligned}\label{eq:crofton1}
		\int_{\Sigma\times \Omega_{d-1}/2}\D \Gamma'_{\Sigma}&=\int _{\Sigma} \D \sigma \int_{\Omega_{d-1}/2} |\cos \alpha_{d} | \D \Omega _{d-1}\\&= \mathrm{Area}(\Sigma) \frac{\Omega_{d-2}}{d-1},
	\end{aligned}
\end{equation}
where the left-hand side counts the total number of intersections (weighted by the measure) on $\Sigma$, while the right-hand side is proportional to the area of $\Sigma$. In summary, we demonstrated that the intrinsic measure \eqref{eq:inv_dens} for intersections on $\Sigma$ implies the following properties:
\begin{itemize}
	\item \textit{At any point on $\Sigma$, the intersectons are isotropically distributed;}
	\item \textit{The distribution of the intersections over $\Sigma$ is uniform.}
\end{itemize}

Alternatively, one can reorganize the same counting by integrating over geodesics rather than over intersections. Let $\mathcal{C}_{\Sigma}$ be the collection of all geodesics that hit $\Sigma$ at least once, and let $\#(\Sigma\cap\Gamma)$ denote the number of intersection points of such a geodesic with $\Sigma$. Then
\begin{equation}\label{eq:crofton2}
	\int_{\Sigma\times \Omega_{d-1}/2}\D \Gamma'_{\Sigma}=\frac12\int_{\mathcal{C}_{\Sigma}}\#(\Sigma\cap \Gamma)\D  \Gamma_{\Sigma},
\end{equation}
where the prefactor $1/2$ corrects for the double counting arising from the fact that every unoriented geodesic contributes two oriented intersections. Combining \eqref{eq:crofton1} and \eqref{eq:crofton2} immediately gives a Crofton-like formula that reconstructs the area of $\Sigma$:
\begin{equation}\label{eq:Crofton}
	\begin{aligned}
		\frac12 \frac{d-1}{\Omega_{d-2}}\int_{\mathcal{C}_{\Sigma}}\#(\Sigma\cap \Gamma)\D  \Gamma_{\Sigma} &= \mathrm{Area}(\Sigma).
	\end{aligned}
\end{equation}
However, in general the set of geodesics $\mathcal{C}_{\Sigma}$ cannot be used to reconstruct the area of other codimension-one surfaces via the Crofton formula.

Now we turn to the special case where $\mathcal{M}$ has a kinematic space $\mathbb{K}_{\mathcal{M}}$ and the area of any codimension-one surface $\Sigma$ can be reconstructed via the Crofton formula. Since the distribution of the geodesics in $\mathbb{K}_{\mathcal{M}}$ is determined by the kinematic measure, we can always read off the measure $\d \Gamma^0_{\Sigma}$ for the intersections between any codimension-one hypersurface $\Sigma$ and the geodesics in $\mathbb{K}_{\mathcal{M}}$. We then demonstrate that this measure $\d \Gamma^0_{\Sigma}$ exactly coincides with the intrinsic measure $\d\Gamma_{\Sigma}'$ of $\Sigma$ given by \eqref{eq:inv_dens}.

Consider an arbitrary point $P$ on $\Sigma$, at which we place an infinitesimal surface element $\sigma$ with an arbitrary normal vector $\textbf{n}$. According to the Crofton formula, the area of $\sigma$ can be reconstructed by counting the number of intersections between $\sigma$ and the geodesics in $\mathbb{K}_{\mathcal{M}}$, which is given by
\begin{align}
	\int_{\sigma \times \Omega_{d-1}/2}\D \Gamma^0_{\sigma} =& \text{Area}(\sigma)\frac{\Omega_{d-2}}{d-1}.
\end{align}
The common factor $\Omega_{d-2}$ and the independence from $\textbf{n}$ indicate that the PEE threads passing through $P$ are isotropically distributed. Furthermore, the fact that the number of intersections is always proportional to the area of $\sigma$ with a universal coefficient implies that the density of intersections for any $\sigma$ is given by the same constant. These properties are identical to those that determine the intrinsic measure $\d\Gamma_{\Sigma}'$, meaning that $\d \Gamma^0_{\Sigma}$ coincides with $\d\Gamma_{\Sigma}'$ up to an overall coefficient. Moreover, they are actually identical to each other since
\begin{align}
	\int_{\Sigma \times \Omega_{d-1}/2}\D \Gamma^0_{\Sigma} = \text{Area}(\Sigma)\frac{\Omega_{d-2}}{d-1}=	\int_{\Sigma\times \Omega_{d-1}/2}\D \Gamma'_{\Sigma}.
\end{align}
These observations lead to the following theorem.
\begin{itemize}
	\item \textit{Theorem 1: For a manifold $\mathcal{M}$ with a kinematic space $\mathbb{K}_{\mathcal{M}}$, the measure of intersections between the geodesics in $\mathbb{K}_{\mathcal{M}}$ and any codimension-one surface $\Sigma$ is always given by the intrinsic measure \eqref{eq:inv_dens}. In other words, we can identify the set of all geodesics in $\mathbb{K}_{\mathcal{M}}$ that pass through $\Sigma$ by shooting geodesics along all intersections on $\Sigma$ with the measure \eqref{eq:inv_dens}.}
\end{itemize}

First, let us reproduce $\mathbb{K}_{\mathcal{M}}$ for vacuum AdS space from the intrinsic measure $d\Gamma_{\Sigma}'$. The key is to find a surface that intersects all geodesics in $\mathbb{K}_{\mathcal{M}}$. Then, according to \textit{Theorem 1}, the set of geodesics shot from the intersections on $\Sigma$ with the intrinsic measure $\d\Gamma_{\Sigma}$ is exactly $\mathbb{K}_{\mathcal{M}}$. In the case of vacuum AdS, such a $\Sigma$ exists and is precisely the AdS boundary. Furthermore, since every geodesic intersects $\Sigma$ at its two endpoints, the measure $\frac{1}{2}\d \Gamma_{\Sigma}'$ should exactly reproduce the kinematic measure $\d\Gamma_{\Sigma}$, where the factor $1/2$ accounts for the fact that one unoriented geodesic corresponds to two intersections.

Now we check this by explicit computation, which has been done in \cite{Lin:2024fze}. For any two boundary points $\textbf{x}_1$ and $\textbf{x}_2$, let $\ell(\textbf{x}_1,\textbf{x}_2)$ denote the length of the bulk geodesic connecting them. This length is obtained by extremizing the action,
\begin{equation}
	\ell(\textbf{x}_1,\textbf{x}_2) = \min_{\textbf{x}_1 \rightarrow  \textbf{x}_2}\int^{\tau_2}_{\tau_1} \D \tau \ \mathcal{L} = \min_{\textbf{x}_1 \rightarrow  \textbf{x}_2}\int^{\tau_2}_{\tau_1}  \D \tau \sqrt{g_{\mu\nu} \dot{\gamma}^\mu \dot{\gamma}^\nu},
\end{equation}
over all curves $\gamma^\mu(\tau)$ interpolating between $\textbf{x}_1$ and $\textbf{x}_2$, where $\textbf{x}_{1,2}=(x_{1,2}^1,x_{1,2}^2, \cdots x_{1,2}^{d-1})$. A standard Hamilton--Jacobi argument identifies the endpoint momentum as the derivative of the on-shell length, $p_\mu|_{\textbf{x}_2}= \partial \ell(\textbf{x}_1,\textbf{x}_2) / \partial x_2^\mu$; hence the momentum data can be traded for the endpoint coordinate $\textbf{x}_2$. The intrinsic measure $\d\Gamma_{\Sigma}'$ \eqref{eq:inv_dens} for the intersections and the corresponding measure for the geodesics $\mathcal{C}_{\Sigma}$ are then given by
\begin{equation}\label{eq:Gamma_ads}
	\D \Gamma_{\Sigma}=\frac{1}{2}\D \Gamma_{\Sigma}' = \frac{1}{2}\det \left( \frac{\partial^2 \ell (\textbf{x}_1,\textbf{x}_2)}{\partial \textbf{x}_1 \partial \textbf{x}_2} \right) \D  \textbf{x}_1 \wedge \D \textbf{x}_2,
\end{equation}
with the shorthand $\D \textbf{x}_{1,2}\equiv\D x^1_{1,2}\wedge\D x^2_{1,2}\wedge\cdots\wedge \D x^{d-1}_{1,2}$ and the prefactor $1/2$ compensating for the fact that each unoriented geodesic contributes two boundary intersections. Specializing to vacuum AdS, where the geodesic length takes the familiar form $\ell(\textbf{x}_1,\textbf{x}_2)=2\log\frac{|\textbf{x}_1-\textbf{x}_2|}{\epsilon}$, a short computation gives
\begin{equation}
	\det \left( \frac{\partial^2 \ell (\textbf{x}_1,\textbf{x}_2)}{\partial \textbf{x}_1 \partial \textbf{x}_2} \right) = \det\left( 2\frac{-\delta ^{ij}|\textbf{x}_{1} -\textbf{x}_{2}|^{2} +2\left( x_{1}^{i} -x_{2}^{i}\right)\left( x_{1}^{j} -x_{2}^{j}\right)}{|\textbf{x}_{1} -\textbf{x}_{2}|^{4}}\right) = \frac{2^{d-1}}{|\textbf{x}_1 - \textbf{x}_2|^{2d -2}}.
\end{equation}
This exactly coincides with the proposal \cite{Czech:2015qta,Czech:2016xec} for the kinematic measure of vacuum AdS, as well as the PEE structure for the vacuum CFT \cite{Lin:2024fze}. 

Next, we generalize the above discussion to derive the kinematic measure for subregions in vacuum AdS. Denote the subregion and its boundary by $\textbf{a}$ and $\partial \textbf{a}$. Here, a geodesic in $\textbf{a}$ means a geodesic chord anchored on $\partial \textbf{a}$, i.e., a portion of a complete geodesic lying within $\textbf{a}$, rather than a full geodesic anchored on the AdS boundary (see Fig.\ref{fig:elementsinKa}). Since every such chord intersects $\partial \textbf{a}$ at its two endpoints, all elements of $\mathbb{K}_{\textbf{a}}$ can be obtained by shooting geodesics inward from $\partial \textbf{a}$. According to \textit{Theorem 1}, shooting geodesics from the intersections on $\partial \textbf{a}$ with the intrinsic measure $\d\Gamma_{\partial\textbf{a}}'$ precisely yields the set of geodesic chords that uniformly cover $\textbf{a}$, namely the kinematic space $\mathbb{K}_{\textbf{a}}$. As in our discussion for the AdS boundary, let $\textbf{s}$ denote the coordinates on $\partial \textbf{a}$, $\ell(\textbf{s}_1,\textbf{s}_2)$ denote the length of the geodesic chord connecting two points on $\partial \textbf{a}$, and $p_\mu |_{\textbf{s}_2}= \partial \ell(\textbf{s}_1,\textbf{s}_2) / \partial s_2 ^\mu$. We thus arrive at the following theorem for the kinematic measure of subregions. 
\begin{itemize}
	\item \textit{Theorem 2: Given a subregion $\textbf{a}$ which has a kinematic space $\mathbb{K}_{\textbf{a}}$, the kinematic measure for $\mathbb{K}_{\textbf{a}}$ can be read from the intrinsic measure \eqref{eq:inv_dens} for the intersections on $\partial \textbf{a}$, which can be written as
		\begin{equation}\label{eq:Gamma_pa}
			\D \Gamma_{\partial \textbf{a}} =\frac{1}{2}\D \Gamma_{\partial \textbf{a}}'= \frac{1}{2}\det \left( \frac{\partial^2 \ell (\textbf{s}_1,\textbf{s}_2)}{\partial \textbf{s}_1 \partial \textbf{s}_2} \right) \D  \textbf{s}_1 \wedge \D \textbf{s}_2,
		\end{equation}
		where $\D \textbf{s}_{1,2}=\D s^1_{1,2}\wedge\D s^2_{1,2}\wedge\cdots\wedge \D s^{d-1}_{1,2}$.}
\end{itemize}
As expected, this reproduces the kinematic measure \eqref{eq:Gamma_a} for subregions in AdS$_3$. Explicit computations for the kinematic measure of subregions will be carried out following this theorem in a separate paper \cite{upcoming}.

Before moving on, let us give a brief summary, which will be useful in our later discussion on subregion reconstruction. 
\begin{enumerate}
	\item \textit{Given a codimension-one hypersurface $\Sigma$ in any manifold, the intrinsic measure \eqref{eq:Gamma_pa} specifies the set of geodesics passing through $\Sigma$ with the density of intersections uniformly distributed over $\Sigma$ and over all directions.}
	\item \textit{If all geodesics in $\textbf{a}$ anchor on $\partial \textbf{a}$, and the kinematic space $\mathbb{K}_{\textbf{a}}$ for the region $\textbf{a}$ exists, then the kinematic measure for $\mathbb{K}_{\textbf{a}}$ is given by \eqref{eq:Gamma_pa}.}
	\item \textit{We can reconstruct the area of any codimension-one surface in $\textbf{a}$ using the Crofton formula \eqref{Crofton} with $\mathbb{K}_{\mathcal{M}}$ replaced by the subregion kinematic space $\mathbb{K}_{\textbf{a}}$, which is a typical instance of subregion reconstruction that we will discuss later.}
\end{enumerate}

\subsection{Subregion reconstruction in AdS}
First, let us define what we mean by reconstruction in this paper. Given a codimension-one surface $\Sigma$, we can determine the set of geodesics $\mathcal{C}_{\Sigma}$ passing through it, together with the intrinsic measure \eqref{eq:inv_dens} for the intersections. If we can find a subregion $\mathcal{W}_{\Sigma}$ in the manifold that is uniformly covered by $\mathcal{C}_{\Sigma}$ in the sense that the area of any codimension-one surface $\gamma$ in $\mathcal{W}_{\Sigma}$ can be reconstructed via the Crofton formula
\begin{equation}\label{Croftonp}
	\text{For any $\gamma$ in the subregion $\mathcal{W}_{\Sigma}$}:~~~~~	\text{Area}\left(\gamma\right)=\frac{1}{2}\frac{d-1}{\Omega_{d-2}} \int_{\mathcal{C}_{\Sigma}}\#\left(\Gamma\cap \gamma\right)d\Gamma.
\end{equation}
Equivalently, for any point in $\mathcal{W}_{\Sigma}$, all the geodesics in $\mathbb{K}$ passing through it belong to $\mathcal{C}_{\Sigma}$. The above subregion reconstruction is purely a geometric statement based on the Crofton formula, which is different from the bulk reconstruction in holographic theories.

The motivation for studying subregion reconstruction is manifold. First, there exist configurations where the reconstruction via the Crofton formula for the entire manifold does not exist, while a subregion reconstruction based on a certain class of geodesics is possible. Second, one can start from a hypersurface $\Sigma$ and identify the subregion that can be reconstructed from the geodesics $\mathcal{C}_{\Sigma}$, as an analogue of holography for subregions in the bulk or the so-called surface-state correspondence. Third, subregion reconstruction provides concrete toy models of the so-called generalized entanglement wedges for gravitational regions \cite{Bousso:2022hlz}.

Before moving on, let us introduce an important concept in geometry: the \textit{totally geodesic submanifold}. 
\begin{itemize}
	\item \textit{A totally geodesic submanifold is a submanifold such that every geodesic of the submanifold (with respect to its induced metric) is also a geodesic of the ambient Riemannian manifold. A submanifold is totally geodesic if and only if its second fundamental form vanishes identically.}
\end{itemize}
Equivalently, a submanifold contains all geodesics of the ambient manifold that start tangent to it. If a geodesic intersects a totally geodesic submanifold, it either meets the submanifold only once or lies entirely within it. Examples of totally geodesic submanifolds include: 1) all geodesics in any Riemannian manifold; 2) equatorial spheres $S^{n-1}$ inside the $n$-dimensional round sphere $S^{n}$; 3) planes in Euclidean spaces, and so on. 

In this paper, we focus on vacuum AdS space; hence the kinematic space $\mathbb{K}_{\textbf{a}}$ exists for all subregions $\textbf{a}$. Moreover, there exist codimension-one totally geodesic submanifolds, namely all hemispheres anchored on the boundary, or the RT surfaces for spherical regions. In the following, we list several cases where subregion reconstruction works.

\subsubsection*{Case 1: Subregions reconstructed by closed convex hypersurfaces}
\begin{figure}[ht]
	\centering
	\includegraphics[width=0.7\linewidth]{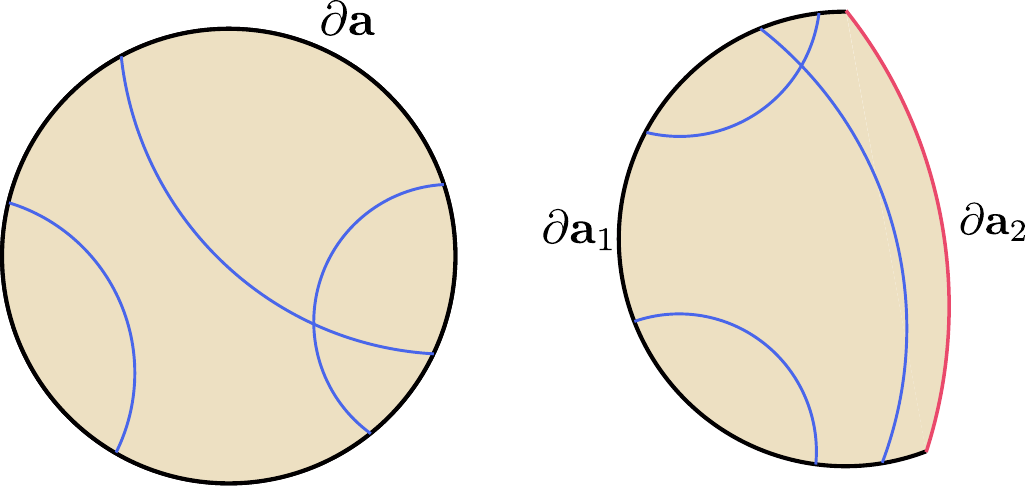}
	\caption{The left figure shows an example of a closed convex surface in AdS, where all geodesics (blue lines) in $\textbf{a}$ belong to $\mathcal{C}_{\textbf{a}}$; hence the region enclosed by $\partial \textbf{a}$ can be reconstructed from $\partial \textbf{a}$. The right figure shows an example of subregion reconstruction from an open convex $\partial \textbf{a}_1$. Here $\partial \textbf{a}_2$ (red line) is a totally geodesic submanifold, so no geodesic can anchor on it. Then all geodesics (blue lines) in $\textbf{a}$ have at least one endpoint anchored on $\partial \textbf{a}_1$, hence they belong to $\mathcal{C}_{\partial \textbf{a}_1}$.}
	\label{fig:case12}
\end{figure}

Let us consider a subregion $\textbf{a}$ enclosed by a closed convex surface $\partial \textbf{a}$; hence all geodesics connecting any two points on $\partial \textbf{a}$ lie within $\textbf{a}$ (see the left figure in Fig.~\ref{fig:case12}). According to our previous discussion, the kinematic space $\mathbb{K}_{\textbf{a}}$ exists and the kinematic measure is given by \eqref{eq:Gamma_pa}. The geodesics in $\mathbb{K}_{\textbf{a}}$ are exactly those shot from $\partial \textbf{a}$, i.e., $\mathcal{C}_{\partial \textbf{a}}$. Then the area of any hypersurface $\gamma$ in $\textbf{a}$ can be reconstructed from $\partial \textbf{a}$ via \eqref{Croftonp}.

Furthermore, since $\partial \textbf{a}$ is convex, for any point $P$ outside $\textbf{a}$, the set of all geodesics passing through $P$ always contains geodesics that do not intersect $\partial \textbf{a}$. This means that any area element outside $\textbf{a}$ cannot be fully reconstructed by the geodesics $\mathcal{C}_{\partial \textbf{a}}$. In summary, the set of geodesics is sufficient to reconstruct the region $\textbf{a}$ and only $\textbf{a}$.

\subsubsection*{Case 2: Subregions reconstructed by open convex hypersurfaces}
Now consider the case where the convex surface is open. The picture is easier to understand in a two-dimensional manifold: we take an open curve $\partial \textbf{a}_1$ with a convex-like property that all geodesics connecting any two points on $\partial \textbf{a}_1$ lie on the same side. At the same time, denote the geodesic connecting the two endpoints of $\partial \textbf{a}_1$ by $\partial \textbf{a}_2$; then $\partial \textbf{a}=\partial \textbf{a}_1\cup\partial \textbf{a}_2$ is a closed curve, and $\textbf{a}$ is the subregion enclosed by $\partial \textbf{a}$ (see the right figure in Fig.~\ref{fig:case12}). Note that $\partial \textbf{a}_2$ is a geodesic, hence a totally geodesic submanifold. This implies that no geodesic in $\textbf{a}$ has both endpoints on $\partial \textbf{a}_2$; consequently, every geodesic in $\textbf{a}$ intersects $\partial \textbf{a}_1$. Then, following our previous discussion, if we set the measure for these geodesics to be \eqref{eq:Gamma_pa}, we obtain the kinematic space $\mathbb{K}_{\textbf{a}}$, which can be used to reconstruct the area of any curve $\gamma$ in $\textbf{a}$ via the Crofton formula \eqref{Croftonp}. Our demonstration for two-dimensional manifolds straightforwardly applies to higher dimensions and leads to the following conclusion.
\begin{itemize}
	\item \textit{Theorem 3: The subregion $\mathcal{W}_{\Sigma}$ that can be reconstructed from $\Sigma$ is the set of all points $P$ that can be enclosed by $\Sigma$ and any other totally geodesic submanifold, such that all geodesics passing through $P$ belong to $\mathcal{C}_{\Sigma}$.}
\end{itemize}
In this theorem, we do not require $\Sigma$ to be convex or open, as we will show in Case 3 that the above theorem works for subregion reconstruction from a generic hypersurface.

A typical example of this reconstruction is to take $\partial\textbf{a}_1$ as a spherical region on the AdS boundary. Since the RT surface $\mathcal{E}_{\textbf{a}_1}$ is a totally geodesic submanifold, all points in the entanglement wedge of $\partial\textbf{a}_1$ belong to the reconstruction region $\mathcal{W}_{\textbf{a}}$. Furthermore, because any point outside the entanglement wedge has at least one geodesic passing through it that is not in $\mathcal{C}_{\partial\textbf{a}_1}$, such points are not in the reconstruction region. In conclusion, the subregion $\mathcal{W}_{\textbf{a}}$ that can be reconstructed by $\partial\textbf{a}_1$ coincides with the entanglement wedge of $\partial\textbf{a}_1$.

One can perform a Rindler transformation \cite{Casini:2011kv}, which maps the entanglement wedge of intervals or spherical regions into the exterior region of the BTZ black brane (or higher-dimensional topological black holes). Then we arrive at the following interesting conclusion.
\begin{itemize}
	\item \textit{Corollary: The region between the horizon of the BTZ black brane (or higher-dimensional topological black holes) and the AdS boundary can be reconstructed from the AdS boundary.}
\end{itemize}
Since the BTZ black hole can be obtained by rolling up the BTZ black brane, there is also a kinematic space $\mathbb{K}_{\textbf{a}}$ for the region $\textbf{a}$ between the AdS boundary $\partial \textbf{a}_1$ and the black hole horizon $\partial\textbf{a}_2$. Here we denote the BTZ black hole horizon as $\partial \textbf{a}_2$ because it is a totally geodesic submanifold. Following our \textit{Theorem 3}, we can reconstruct the BTZ black hole exterior region from the AdS boundary.

Now consider $\partial \textbf{a}_2$ to be the extremal surface that anchors on the boundary of $\partial \textbf{a}_1$, which more closely resembles the entanglement wedge. However, in this case $\partial \textbf{a}_2$ is generally not a totally geodesic submanifold, so there exist geodesics in $\textbf{a}$ that do not intersect $\partial \textbf{a}_1$. For example, if we consider the RT surface for a strip region in AdS$_4$, we can find geodesics in the entanglement wedge whose endpoints both lie on the RT surface. This means that not all points can be reconstructed from $\mathcal{C}_{\partial\textbf{a}_1}$ alone. In conclusion, the region enclosed by $\partial \textbf{a}_1 \cup \partial \textbf{a}_2$ can be reconstructed from $\partial \textbf{a}_1 \cup \partial \textbf{a}_2$, but not from $\partial \textbf{a}_1$ alone. For boundary regions $\partial \textbf{a}_1$ that are not spherical, the entanglement wedge no longer coincides with $\mathcal{W}_{\textbf{a}}$.

\begin{figure}[ht]
	\centering
	\includegraphics[width=0.7\linewidth]{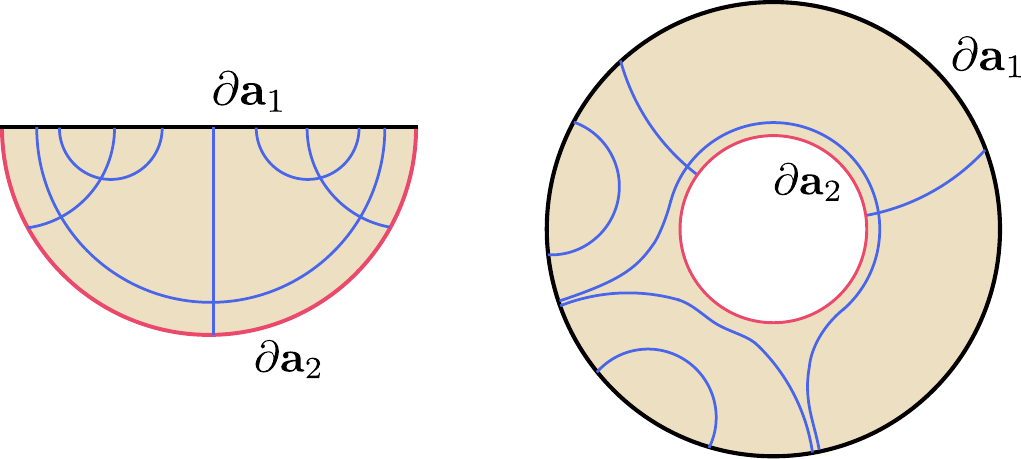}
	\caption{The left figure shows an example of subregion reconstruction from a boundary interval $\partial \textbf{a}_1$. The right figure shows the reconstruction of the exterior of the BTZ black hole from the asymptotic AdS boundary. In both cases, $\partial \textbf{a}_2$ is a totally geodesic submanifold, and all geodesics in the region enclosed by $\partial \textbf{a}_1$ and $\partial \textbf{a}_2$ are shot from $\partial \textbf{a}_1$.}
	\label{fig:casebtz}
\end{figure}

\subsubsection*{Case 3: Subregions reconstructed by non-convex hypersurfaces}
\begin{figure}[ht]
	\centering
	\includegraphics[width=0.55\linewidth]{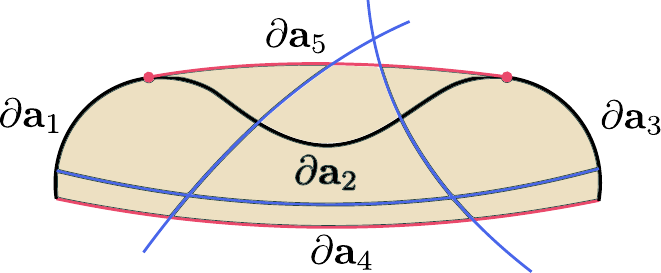}
	\caption{Illustration of the subregion reconstruction from a non-convex curve.}
	\label{fig:noncovex}
\end{figure}

We discuss non-convex hypersurfaces separately because they can reconstruct regions not only from one side but also from the other side. For example, consider a curve $\pab_1\cup\pab_2\cup \pab_3$ in AdS that can be divided into three convex parts, such that all geodesics connecting points in $\pab_1$ lie below it, and similarly for $\pab_3$, while all geodesics connecting points in $\pab_2$ lie above it. We can draw two geodesics $\pab_4$ and $\pab_5$, which are totally geodesic submanifolds, connecting the endpoints of $\pab_1\cup\pab_2\cup \pab_3$ and of $\pab_2$, respectively (see Fig.~\ref{fig:noncovex}). According to our previous discussion, the region enclosed by $\pab_2$ and $\pab_4$ can be reconstructed from $\pab_2$. Furthermore, the curve $\pab_1\cup\pab_5\cup\pab_3$ is now convex, so by \textit{Theorem 3} it can reconstruct the region $\ab$ enclosed by $\pab_1\cup\pab_4\cup\pab_3\cup \pab_5$. In conclusion, the maximal region $\ab$ that can be reconstructed from the non-convex curve $\pab_1\cup\pab_2\cup \pab_3$ lies on both sides of the curve.

Our discussion for the case in Fig.~\ref{fig:noncovex} can be extended to subregion reconstruction from more complicated non-convex curves and from higher-dimensional non-convex hypersurfaces. The exact region that can be reconstructed from a generic hypersurface in AdS is still described by our \textit{Theorem 3}, i.e., the set of all points that can be enclosed by this hypersurface together with any totally geodesic submanifold.

\section{Surface-state correspondence realized in the PEE tensor network models}
\label{sec3}
Our subregion reconstruction from hypersurfaces sheds light on constructing toy models of holography for gravitational subregions, which is the duality between the quantum states on the hypersurface and the states of the subregions. Following \cite{Wen:2025gui}, we will build tensor network models for subregions that can be reconstructed from a hypersurface, as holographic toy models for the subregion holography. Similar properties in the so-called surface/state correspondence conjecture \cite{Miyaji:2015yva} (see also \cite{Miyaji:2015fia,Chen:2019mis,Bao:2023til} for relevant developments) can be reproduced in our toy models.

\subsection{The surface/state correspondence}
\begin{figure}[ht]
	\centering
	\includegraphics[width=0.65\linewidth]{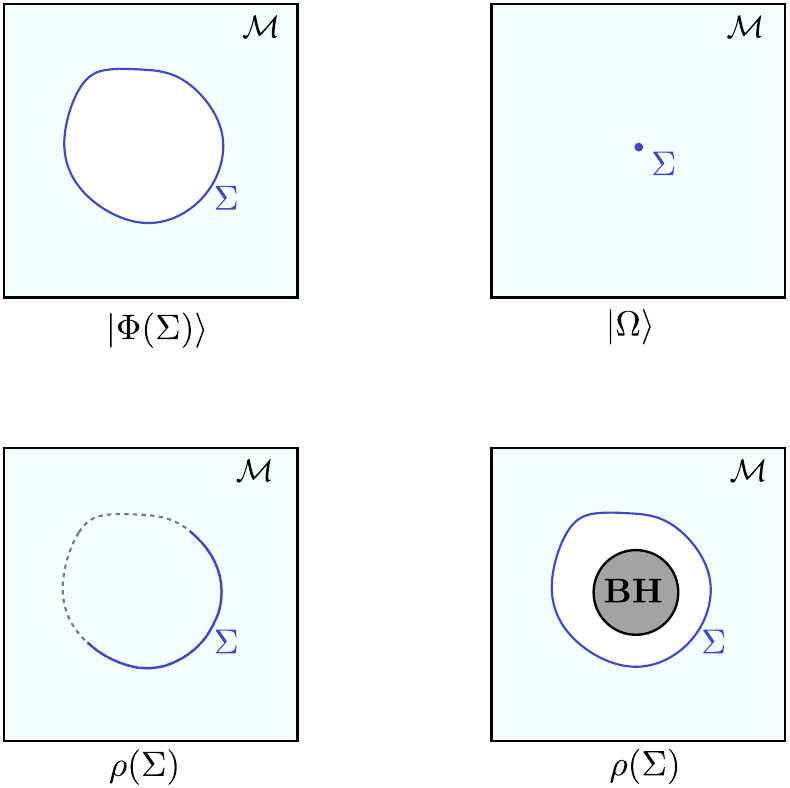}
	\caption{Illustration of the surface/state correspondence. A closed convex surface $\Sigma$ that is topologically trivial corresponds to a pure state $\ket{\Phi(\Sigma)}$. An open convex surface or a topologically non-trivial closed surface corresponds to a mixed state $\rho({\Sigma})$.}
	\label{fig:rt-ss}
\end{figure}

Let us briefly introduce the surface/state correspondence, which is a conjectured generalization of the standard AdS/CFT correspondence. While standard holography equates a bulk gravity theory to a quantum field theory on the AdS boundary, the surface/state correspondence proposes a much more radical, local relation. Consider a gravitational theory on a bulk manifold $\CM$ (now including the time direction), associated with a large Hilbert space $\mathcal{H}_{\CM}$. Take a convex, codimension-2 surface $\Sigma$ in $\CM$. The surface/state correspondence mainly contains the following relations (see Fig.\ref{fig:rt-ss}) \cite{Miyaji:2015yva}:
\begin{enumerate}
	\item A closed convex surface $\Sigma$ that is topologically trivial, i.e., homologous to a point, corresponds to a pure state $\ket{\Phi(\Sigma)}$ in $\mathcal{H}$. Taking the zero-size limit of this pure-state leads to the state $\ket{\Omega}$ with no real-space entanglement,
	\begin{align}
		\lim_{{\rm Area}(\Sigma)\to 0}\ket{\Phi(\Sigma)}=\ket{\Omega}\,.
	\end{align}
	\item A closed $\Sigma$ that is topologically non-trivial, for example a surface surrounding a black hole, corresponds to a mixed state $\rho(\Sigma)$ in the Hilbert space $\mathcal{H}_{\Sigma}$, which is a subspace of $\mathcal{H}$.
	\item An open surface $\Sigma$ also corresponds to a mixed state $\rho(\Sigma)$ in $\mathcal{H}_{\Sigma}$, which can be understood as the reduced density matrix of a subregion in the pure state of a closed convex surface.\footnote{When $\Sigma$ is a subregion in a closed convex surface $\tilde{\Sigma}$, $\Sigma$ corresponds to the mixed state obtained by tracing out the degrees of freedom in the complement of $\Sigma$ within $\tilde{\Sigma}$.}
	\item A smooth deformation preserving convexity from a surface $\Sigma_1$ to another surface $\Sigma_2$ corresponds to a series of infinitesimal unitary transformations. Given two convex topologically trivial surfaces $\Sigma_{1}$ and $\Sigma_2$, we have
	\begin{align}
		\ket{\Psi(\Sigma_1)}=U(s_1,s_2)\ket{\Psi(\Sigma_2)},\qquad U(s_1,s_2)=P \cdot \exp\left(-i\int_{s_1}^{s_2}M(s)\,ds\right),
	\end{align}
	where $P$ denotes path ordering, $M(s)$ is a Hermitian operator, and the parameter $s$ characterizes the continuous deformation, with $s=s_{1,2}$ corresponding to the surfaces $\Sigma_{1,2}$. 
	
	\item When $\Sigma_{1,2}$ are two open surfaces with the same boundary $\partial\Sigma_1=\partial\Sigma_2$, the density matrices are related by a unitary transformation:
	\begin{align}
		\rho(\Sigma_2)=U(s_1,s_2)^{-1}\rho(\Sigma_1)U(s_1,s_2).
	\end{align}
\end{enumerate} 

\begin{figure}[ht]
	\centering
	\includegraphics[width=0.3\textwidth]{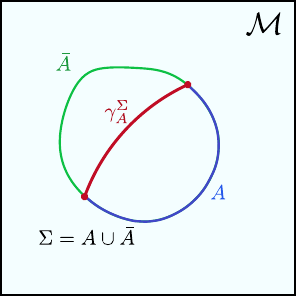}
	\caption{The RT-like prescription for entanglement entropy in the surface/state correspondence}
	\label{fig:ss2}
\end{figure}

In the context of AdS/CFT and considering the vacuum AdS, the gravitational ``sandwich'' between a topologically trivial (or closed) surface $\Sigma$ and the asymptotic boundary (or between two surfaces) is dual to a linear map (a quantum channel) that prepares the state $\ket{\Phi(\Sigma)}$ as a unitary evolution of the boundary state. In other words, one can cut spacetime anywhere along a closed surface, and that cut defines a quantum state on the surface. The entanglement entropy for subsystems on $\Sigma$ can be computed by an analogue of the RT surface, which is the extremal surface in the region homologous to the subsystem. For example, consider a subregion $A$ on the closed convex surface $\Sigma$ (see Fig.\ref{fig:ss2}). The RT-like proposal for the entanglement entropy is given by
\begin{align}
	S_A^\Sigma=\frac{\text{Area}(\gamma^\Sigma_A)}{4G}.
\end{align}
\subsection{The PEE tensor network}
\begin{figure}[ht]
	\centering
	\includegraphics[width=0.5\linewidth]{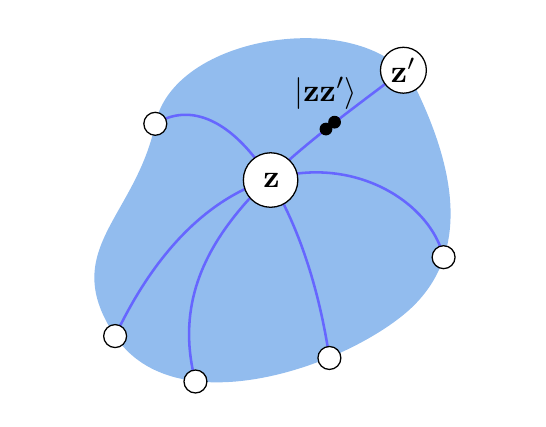}
	\caption{This figure is extracted from \cite{Wen:2025gui}. A tensor state $\ket{\T(\v{z})}$ located at the bulk site $\v{z}$ (hollow white dot). The legs emanating from $\v{z}$ are continuously distributed and can be described by a vector field as in \cite{Lin:2023rxc}. The contraction between tensors at $\v{z}$ and an adjacent site $\v{z'}$ is performed by projecting onto a maximally entangled state $\ket{\v{z}\v{z'}}$ along the PEE thread connecting them.}
	\label{fig:PEE network}
\end{figure}

Now we introduce a new type of tensor network models for AdS/CFT, called the PEE tensor networks \cite{Wen:2025gui}. These models are characterized by a network structure formed by geodesics in Kinematic space, which we refer to as PEE threads. For any point on the AdS boundary, we can use a vector field to describe all the PEE threads emanating from that point: the norm of the vector field captures the density of the threads, and the threads themselves are the integral curves of the vector field. The full PEE network is then constructed by superposing all such vector fields (see an example for $d=2$ in Fig.\ref{fig:PEE network}). The resulting PEE network provides a perfect tessellation of the vacuum AdS geometry. Note that the number of intersections is computed by the ``flux'' of the vector fields through a given surface. Since the PEE threads are unoriented, here "flux" means the total number of crossings, rather than the net flux where contributions from ingoing and outgoing threads cancel.

In usual discrete tensor network models, the entanglement entropy is typically given by the number of intersections between a surface and the network. In AdS/CFT, by contrast, the entanglement entropy is captured by the area of the Ryu–Takayanagi (RT) surface in a continuous geometric background. Following the Crofton formula, counting the number of intersections in the PEE network exactly corresponds to computing the area of a smooth surface. In other words, the PEE tensor network models can precisely reproduce the RT formula, and therefore should be regarded as a more accurate tensor network model of AdS/CFT.

Like discrete tensor networks, we use a local tensor $\T_{u_1u_2\cdots} (\v{z})$ to represent a quantum state $\ket{\T(\v{z})}$ located at site $\v{z}$,
\begin{equation}\label{vertexstate1}
	\ket{\T(\v{z})} \equiv \T_{u_1u_2\cdots} (\v{z}) \ket{u_1}_{\v{z}}\ket{u_2}_{\v{z}}\cdots=\T_{\v{u}}(\v{z})\left(\bigotimes_{i}\ket{u_{i}}_{\v{z}}\right).
\end{equation}
where the index $u_i$ characterizes the leg that is associated with a Hilbert space $\mathcal{H}_{i}$ spanned by the orthonormal basis vectors $\{\ket{u_i}_{\v{z}}\}$, and $\v{u}$ represents the collection of all the $u_i$ indices. Note that in the PEE network the legs are continuously distributed; hence $i$ is a continuous parameter.

For any pair of adjacent sites $\v{z}$ and $\v{z'}$, there is a unique PEE thread connecting them, which picks a pair of legs associated with $\v{z}$ and $\v{z'}$ respectively (see Fig.\ref{fig:PEE network}). Again, as in the discrete models, we contract all such pairs of legs by projecting onto a maximally entangled state to glue adjacent sites. Eventually we arrive at the PEE tensor network, which represents a quantum state $\ket{\Psi}$ for the uncontracted boundary legs,
\begin{equation}\label{contraction0}
	\ket{\Psi}=	\left(\bigotimes_{\v{z},\v{z'} \in \mathcal{M}} \bra{\v{z}\v{z'}}\right)\left(\bigotimes_{\v{z}\in \mathcal{M}} \ket{\T(\v{z})}\right)
\end{equation}
where $ \ket{\v{z}\v{z'}}=\frac{1}{\sqrt{v}}\sum_{k=1}^{k=v}\ket{k}_{\v{z}}\ket{k}_{\v{z'}}$ is a maximally entangled state \cite{Verstraete:2003wvw} defined on any pair of legs connecting the nearby sites $\v{z}$ and $\v{z'}$. This way of contraction can be understood as the gluing of nearby geometries through the imposition of identical boundary conditions in the gravitational path integral.

Then we insert explicit quantum states on the bulk sites. In \cite{Wen:2025gui}, three types of PEE tensor network models are constructed, including:
\begin{enumerate}

	\item \textit{the factorized PEE tensor network constructed by putting tensor product states of EPR pairs on each bulk site; }
	\item \textit{the HaPPY-like PEE tensor network constructed by putting perfect tensors on each bulk site;}
	\item \textit{the random PEE tensor network constructed by putting random states on each bulk site.}
\end{enumerate}
In the first two models, a straightforward computation of the entanglement entropy for spherical regions exactly reproduces the Ryu-Takayanagi (RT) formula, whereas the third random PEE tensor network model reproduces it for generic boundary regions.

In what follows, we focus mainly on the factorized PEE tensor network model, which is the simplest one. In this model, the quantum state at each site $\textbf{z}$ is a tensor product of EPR pairs, where each EPR pair is defined on a pair of legs that belong to the same PEE thread passing through $\textbf{z}$. The local tensor can be written as
\begin{equation}\label{eq:bulk local tensor}
	\ket{\T(\v{z})}=\bigotimes_{\{i,j\}}\T_{ab}(\v{z})\ket{a}_{\v{z}i}\ket{b}_{\v{z}j}\,,
\end{equation}
where $\{i,j\}$ labels any pair of legs lying on the same PEE thread, and the Einstein summation convention is assumed for repeated dummy indices. Since $\T_{ab}$ represents a maximally entangled EPR state, we have $\T^{\dagger}_{ab}\T_{bc}=\frac{1}{v}\delta_{ac}$, or more abstractly $\T^\dagger\T\propto\mathbb{I}$. The contraction of any two neighboring EPR pairs on the same PEE thread then results in the following state,
\begin{align}
&\frac{1}{\sqrt{v}}\sum_{a,b,c,d,e}\bra{e}_{\v{z}}\bra{e}_{\v{z'}}	\T^{1}_{ab}\T^{2}_{cd} \ket{a}_{\v{z}}\ket{b}_{\v{z}}\ket{c}_{\v{z'}}\ket{d}_{\v{z'}} 
\cr
=& \frac{1}{\sqrt{v}} \sum_{a,d,e} \T^{1}_{ae}\T^{2}_{ed}\ket{a}_{\v{z}}\ket{d}_{\v{z}'}
\cr
=& \frac{1}{\sqrt{v}} \sum_{a,d} \T^{3}_{ad}\ket{a}_{\v{z}}\ket{d}_{\v{z}'}\,,
\end{align}
where $\T^{3}=\T^{1}\T^{2}$, and we have omitted the $\{i,j\}$ indices. Since $\T^{1\dagger}\T^{1}\propto \mathbb{I}$ and $\T^{2\dagger}\T^{2}\propto \mathbb{I}$, we have $\T^{3\dagger}\T^{3}= \T^{2\dagger}\T^{1\dagger}\T^{1}\T^{2}\propto \mathbb{I}$, which represents an EPR pair supported on the geodesic chord that contains $\v{z}$ and $\v{z}'$. Contracting all the EPR pairs along a given PEE thread iteratively in this manner yields a single EPR pair associated with this thread.

Next, we contract all the internal legs to get the tensor network, which is just a tensor product state of EPR pairs associated with each PEE thread (see \cite{Wen:2025gui} for details)
\begin{align}\label{boundarystate}
	\ket{\Psi}=\bigotimes_{_{\mathcal{C}(\v{x},\v{y})}} T_{ab}(\v{x},\v{y})\ket{a}_{\v{x}}\ket{b}_{\v{y}}\,.
\end{align}
Here $T_{ab}(\v{x},\v{y})\ket{a}_{\v{x}}\ket{b}_{\v{y}}$ is an EPR state supported on a pair of legs associated with the endpoints of a geodesic $\mathcal{C}(\v{x},\v{y})$ connecting the boundary points $\v{x}$ and $\v{y}$. This pair of legs is tangent to $\mathcal{C}(\v{x},\v{y})$.

\subsection{Realization of the Surface/state correspondence in the factorized PEE tensor network}
\begin{figure}[ht]
	\centering
	\includegraphics[width=0.7\linewidth]{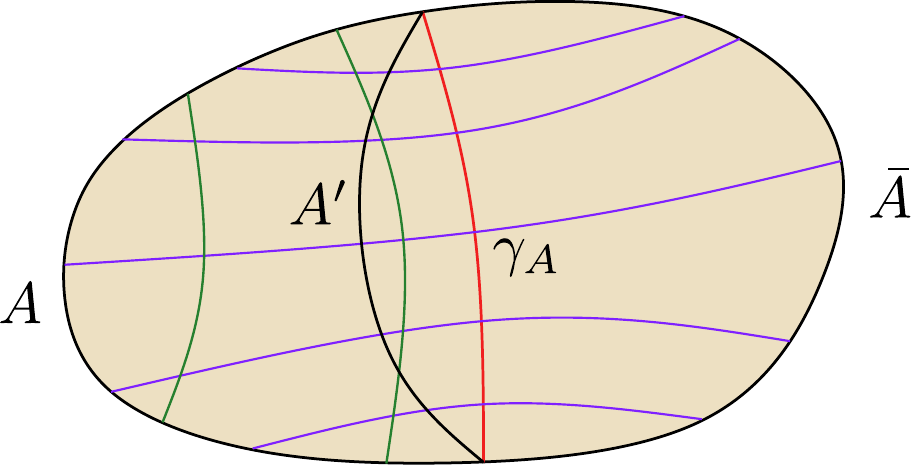}
	\caption{Here $A\cup\bar{A}$ is a closed curve on a time slice of vacuum AdS$_3$, and the red curve $\gamma_A$ is the geodesic homologous to $A$. The blue curves are the PEE threads connecting $A$ and $\bar{A}$, representing the outer entangled EPR pairs, while the green curves are the PEE threads connecting points within $A$, representing the inner entangled EPR pairs. }
	\label{fig:SSinPEETN}
\end{figure}

Now we build tensor network models for sub-regions in AdS, and check that all the properties proposed by the surface/state correspondence can be realized in these models. 

\textbf{A topologically trivial closed surface corresponds to a pure state:} Let us consider a sub-region $\textbf{a}$ in vacuum AdS enclosed by a convex closed surface $A$. The sub-region is covered by the factorized PEE tensor network we just defined. After contracting all the legs inside $\textbf{a}$, we obtain a tensor network which represents a product state of EPR pairs, which is in one-to-one correspondence with the PEE threads in $\textbf{a}$. Since the tensor network represents a pure state on the boundary $A$ curve, we confirmed the statement in the surface/state correspondence that, a closed convex surface that is topologically trivial corresponds to a pure state. One can also consider non-convex closed curves, and the corresponding state represented by our PEE tensor network is also a product state of EPR pairs represented by all the PEE threads, which is also a pure state. More explicitly, the state on the surface is given by
\begin{align}\label{boundarystate1}
	\ket{\Phi(A)}=\bigotimes_{_{\mathcal{C}(\v{s}_1,\v{s}_2)}} T_{ab}(\v{s}_1,\v{s}_2)\ket{a}_{\v{s}_1}\ket{b}_{\v{s}_2}\,,
\end{align}
where $\v{s}_1$ and $\v{s}_2$ represent two sites on $A$ and $\mathcal{C}(\v{s}_1,\v{s}_2)$ is the PEE thread connecting the two sites. 

If we shrink $A$ to be zero size such that there is only one bulk site $\v{z}$ in the surface, then the state on $A$  is exactly the un-contracted tensor state $\ket{\T(\v{z})} $ \eqref{eq:bulk local tensor} on this site, which is also a pure state.

\textbf{An open surface corresponds to a mixed state:} Then we consider an open surface $A$, which can always be considered as a portion of a closed surface $A\cup \bar{A}$, which corresponds to a pure state $\ket{\Phi(A\cup\bar{A})}$ according to our previous discussion (see Fig.\ref{fig:SSinPEETN}). In this case, the legs in $A$ are in a mixed state, given precisely by the reduced density matrix obtained by tracing out the legs in the $\bar{A}$ region. The legs in $A$ may be entangled either among themselves or with legs in $\bar{A}$. These two types of entanglement correspond, respectively, to geodesics with both endpoints in $A$ (green lines), and geodesics with one endpoint in $A$ and the other in $\bar{A}$ (blue lines). Clearly, only the legs entangled with legs in $\bar{A}$ contribute to the entanglement entropy $S_A$, which can be computed by counting the number of EPR pairs, or equivalently the number of PEE threads connecting $A$ and $\bar{A}$. Now consider the geodesic $\gamma_A$ (red line) that is homologous to $A$. Since $\gamma_A$ is a totally geodesic submanifold, we have the following two properties: (1) every PEE thread that intersects $\gamma_A$ connects a point in $A$ to a point in $\bar{A}$; (2) each such PEE thread intersects $\gamma_A$ exactly once. It then follows that the PEE threads connecting $A$ and $\bar{A}$ are precisely those that pass through $\gamma_A$, and the number of these threads is captured by the area of $\gamma_A$ according to the Crofton formula. Note that the measure for the PEE threads passing through $\gamma_A$ is given by the intrinsic measure $\Gamma'_{\gamma_A}$ \eqref{eq:inv_dens} up to an overall coefficient. To match the RT formula, we set this coefficient to $\frac{1}{4G}$ and require each thread to contribute one unit to $S_A$. Hence we have
\begin{align}
	S_A =& \frac{Area(\gamma_A)}{4G},
\end{align}
which is exactly the property proposed by the surface/state correspondence.

\textbf{Homologous open surfaces are related by a unitary transformation:} Now we deform the open curve $A$ to a homologous surface $A'$ while preserving convexity (see also Fig.\ref{fig:SSinPEETN}). It is clear that our analysis for the entanglement entropy $S_A$ also applies to $A'$, hence
\begin{align}
	S_{A'} =& \frac{Area(\gamma_A)}{4G},
\end{align}
which also coincides with the property of the surface/state correspondence. 

However, in the surface/state correspondence conjecture \cite{Miyaji:2015yva}, the states on $A$ and $A'$ are related by a unitary transformation. In contrast, in our model the Hilbert spaces for $A$ and $A'$ are different, since they have different areas and hence different numbers of legs. We can extend the PEE threads enclosed by $A'\cup \bar{A}$ to those enclosed by $A\cup\bar{A}$ by contraction and make the following observations.
\begin{enumerate}
	\item Each leg in $A'$ that is entangled with a leg on $\bar{A}$ corresponds to one and only one leg in $A$, which is entangled to the same leg on $\bar{A}$ and lies on the extension of the PEE thread connecting this entangled pair on $A'$ and $\bar{A}$. See the blue threads in Fig.\ref{fig:SSinPEETN}.
	\item All the inner EPR pairs in $A'$ are represented by PEE threads connecting two points on $A'$, and each of them can be extended to a PEE thread connecting two points on $A$, hence corresponding to one and only one inner EPR pair in $A$. See the right green thread in Fig.\ref{fig:SSinPEETN}.
	\item There are PEE threads connecting two points in $A$ that do not intersect $A'$ (see the left green thread in Fig.\ref{fig:SSinPEETN}), and hence there are extra inner EPR pairs in $A$ that have no counterpart in $A'$.
\end{enumerate}
In summary, the state on $A'$ can be considered as a unitary evolution (doing contraction along the geodesics) of the outer entangled legs and part of the inner entangled legs of $A$. The remaining inner entangled legs are the extra inner entangled EPR pairs which gradually disappear during the deformation from $A$ to $A'$, and they do not contribute to the entanglement entropy. The surface $\gamma_A$ is an extremal case whose inner EPR pairs all disappear during the deformation from $A$ to $\gamma_A$.

\textbf{A topologically non-trivial closed surface corresponds to a mixed state:}
\begin{figure}[ht]
	\centering
	\includegraphics[width=0.5\linewidth]{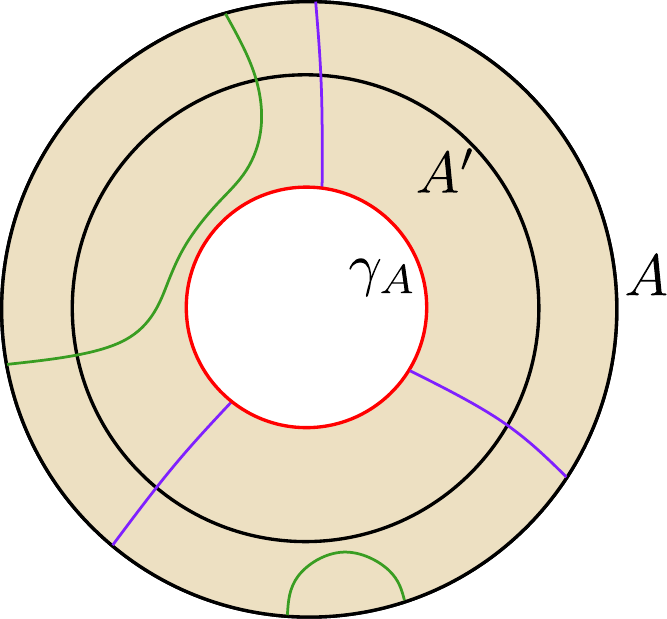}
	\caption{Here, $A$ and $\gamma_A$ denote the AdS boundary and the horizon of the BTZ black hole, respectively. $A'$ is a closed surface enclosing the horizon. The blue curves are the PEE threads connecting $A$ and $\gamma_A$, and they contribute to the entropy of the black hole, while the green curves are the PEE threads representing the inner EPR pairs within $A$.}
	\label{fig:SSBTZ}
\end{figure}
Now we consider the BTZ black hole, and denote the boundary and the horizon by $A$ and $\gamma_A$, respectively (see Fig. \ref{fig:SSBTZ}). According to our previous discussion, the PEE threads emanating from $A$ reconstruct the region between $A$ and $\gamma_A$. Then we can construct a PEE tensor network in this region; after contracting all the internal legs inside this region, we obtain a pure state for all the legs on $A\cup \gamma_A$. The PEE threads anchored on $A$ represent the inner EPR pairs on $A$, while the PEE threads connecting $A$ and $\gamma_A$ represent the EPR pairs with one qudit on $A$ and the other on $\gamma_A$. Note that the horizon is a totally geodesic submanifold; hence there are no inner EPR pairs within $\gamma_A$. By tracing over the legs on $\gamma_A$, we obtain a mixed state for $A$. The entanglement entropy $S_A$ can then be computed by counting the EPR pairs connecting $A$ and $\gamma_A$, which is proportional to the area of $\gamma_A$,
\begin{align}
	S_A= &  \frac{Area(\gamma_A)}{4G}.
\end{align}
Similarly, we can deform $A$ to some closed curve $A'$ in the bulk surrounding the black hole (see Fig. \ref{fig:SSBTZ}), which is the topologically non-trivial closed surface defined in \cite{Miyaji:2015yva}. Again, the state on $A'$ can be regarded as the unitary evolution of the outer entangled legs and part of the inner entangled legs on $A$ (legs connected by a PEE thread which also intersects $A'$). The remaining inner entangled legs on $A$ are the inner EPR pairs represented by the PEE threads anchored on $A$ that do not intersect $A'$ (for example, the legs connected by the small green PEE thread in Fig. \ref{fig:SSBTZ}); these disappear gradually as we deform $A$ to $A'$. So the state on $A'$ is also a mixed state, and $S_A'=S_A$. These properties are also consistent with those proposed for the surface/state correspondence.

\textbf{In summary}: We have reproduced nearly all the properties proposed by the surface/state correspondence for the factorized PEE tensor network model on AdS$_3$. A similar discussion can be extended to higher-dimensional surfaces whenever the extremal surface $\gamma_A$ can be replaced by a totally geodesic submanifold, i.e., (any portion of) the hemispheres anchored on the boundary. Furthermore, our discussion can also be generalized to HaPPY-like tensor networks on AdS$_3$, where the unitary transformations from one surface to another can be implemented in a concrete manner.

\section{A realization of the generalized entanglement wedges for gravitational regions}
\label{sec4}

In AdS/CFT, subregion–subregion duality was originally formulated for boundary subregions and their associated entanglement wedges in the bulk. In \cite{Bousso:2022hlz} (see also \cite{Bousso:2023sya,Kaya:2025vof,Arayath:2026rll,Geng:2020fxl,Geng:2023qwm} for related developments), the authors further conjectured that gravitational regions $\ab$ can likewise be assigned a \emph{generalized entanglement wedge} $\mathcal{W}(\ab)$. More explicitly, consider a static slice $\Sigma$ in AdS and a subregion $\ab \subset \Sigma$. The generalized entanglement wedge $\mathcal{W}(\ab)$ is defined as a gravitational region satisfying
\begin{align}
	\ab\subset \mathcal{W}(\ab)\subset \Sigma,
\end{align}  
and having the smallest generalized entropy among all possible choices that meet the above requirement. Here, the generalized entropy of a gravitational region $\CW$ is
\begin{align}
	S_{gen}(\mathcal{W}) =\frac{\text{Area}(\partial \mathcal{W})}{4G}+S(\mathcal{W})+\cdots, \qquad S(\mathcal{W})=-\tr \rho_{\CW}\log \rho_{\CW}\,,
\end{align}
where $\rho_{\CW}$ is the reduced density matrix of the quantum fields restricted to $\CW$, and $S(\mathcal{W})$ is the bulk entanglement entropy of $\CW$. Additionally, if the boundary $\partial\ab$ of $\ab$ contains a subset $\tilde{\partial} \ab$ lying on the AdS boundary, we require that $\partial\CW(\ab)$ shares the same boundary subset, i.e. $\tilde{\partial}\ab=\tilde{\partial}\CW(\ab)$. When $\ab$ itself is a subregion on the AdS boundary, the generalized entanglement wedge reduces to the standard entanglement wedge enclosed by $\ab$ and its corresponding RT surface $\gamma_{\ab}$.

\begin{figure}[ht]
	\centering
	\includegraphics[width=1\linewidth]{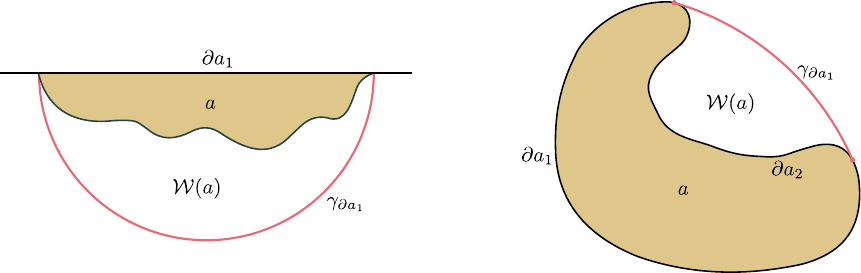}
	\caption{Left: a bulk region $\ab$ whose boundary partially overlaps with the AdS boundary $\partial\ab_1$. The generalized entanglement wedge $\CW(\ab)$ coincides with the entanglement wedge of $\partial\ab_1$. Right: a bulk region $\ab$ deep inside the AdS bulk with no overlap with the AdS boundary. The generalized entanglement wedge $\CW(\ab)$ is bounded by $\partial\ab_1\cup\gamma_{\partial\ab_1}$, where $\gamma_{\partial\ab_1}$ is tangent to $\partial\ab$ at the intersection points.}
	\label{fig:GW12}
\end{figure}

The generalized entanglement wedge is inspired by the configurations where the entanglement wedge of the Hawking radiation of an evaporating black hole includes an additional island region, as well as the tensor network models for static states in quantum gravity. It was discussed in \cite{Bousso:2022hlz} that the generalized entanglement wedge also satisfies a no-cloning theorem, appropriate forms of strong subadditivity and nesting.

Since $\CW(\ab)$ is generally larger than $\ab$, and inspired by the entanglement reconstruction program in AdS/CFT, the generalized entanglement wedge conjecture proposes that the region inside $\CW(\ab)$ but outside $\ab$ can be reconstructed from the information contained in $\ab$. We now attempt to realize this statement using our subregion reconstruction based on the subregion kinematic space.

In the left panel of Fig.~\ref{fig:GW12}, we consider a bulk region $\ab$ whose boundary includes an interval $\partial\ab_1$ on the AdS boundary. Furthermore, we require $\ab$ to be restricted to the entanglement wedge of $\partial\ab_1$, i.e. the bulk region enclosed by the union of $\partial\ab_1$ and its corresponding RT surface $\gamma_{\partial\ab_1}$. In this case, the generalized entanglement wedge $\CW(\ab)$ is exactly the entanglement wedge of $\partial\ab_1$. Note that, as long as $\partial\ab_1$ is fixed, for any region $\ab$ that lies inside the entanglement wedge of $\partial\ab_1$, the generalized entanglement wedge $\CW(\ab)$ is always given by the entanglement wedge of $\partial\ab_1$. Moreover, the minimal case occurs when $\ab$ reduces to the boundary interval $\partial\ab_1$ (viewed as a boundary-localized region), for which $\CW(\ab)$ coincides with the entanglement wedge of $\partial\ab_1$. This matches our subregion reconstruction shown in the left panel of Fig.~\ref{fig:casebtz}, where the entanglement wedge can be reconstructed from the PEE threads emanating from the boundary interval $\partial\ab_1$.

In the right panel of Fig.~\ref{fig:GW12}, we consider a bulk region $\ab$ whose boundary has no overlap with the AdS boundary, so $\ab$ is deep inside the bulk. According to the generalized entanglement wedge conjecture, in this case $\CW(\ab)$ is larger than $\ab$, with $\gamma_{\partial\ab_1}\cup \partial \ab_1$ forming the boundary of $\CW(\ab)$, where $\gamma_{\partial \ab_1}$ is the geodesic homologous to $\partial \ab_1$. Here we take the partition $\partial \ab =\partial \ab_1\cup \partial\ab_2$, with $\partial \ab_1$ a convex curve. If we ignore the bulk entanglement entropy, then minimizing the generalized entropy means minimizing the area of $\partial\ab_1\cup\gamma_{\partial\ab_1}$, hence the geodesic $\gamma_{\partial \ab_1}$ should be tangent to $\partial\ab$ at the intersection points. In this situation, we can vary $\partial \ab_2$ while keeping $\partial \ab_1$ fixed, such that $\CW(\ab)$ remains the region bounded by $\partial\ab_1\cup \gamma_{\partial \ab_1}$. The minimal region $\ab$ needed to reconstruct $\CW(\ab)$ is, in the limiting sense, the surface $\partial \ab_1$ itself, so that $\CW(\ab)=\CW(\partial\ab_1)$. This coincides with our subregion reconstruction shown in the right panel of Fig.~\ref{fig:case12}, which states that the subregion reconstructable from the PEE threads emanating from $\partial \ab_1$ is precisely $\CW(\partial \ab_1)$.

\section{Discussion}
\label{sec5}
In this paper, we introduce a reconstruction scheme for a subregion from a codimension‑one surface in a Riemannian manifold, using kinematic space. The reconstruction is purely geometric: the subregion can be perfectly covered by the PEE threads emanating from the hypersurface. In other words, for any surface inside the subregion, all geodesics passing through it belong to these PEE threads, and the area of that surface can be reconstructed by counting the number of intersections with the threads. We focus on the vacuum AdS case, derive the kinematic measure for arbitrary subregions, and analyze how the subregion reconstruction works based on the PEE threads emitted from an arbitrary codimension‑one surface.

In \cite{Wen:2025gui}, tensor network toy models based on the network of PEE threads have been developed as a simulation of AdS/CFT. Since counting the intersections between a surface and the PEE network is equivalent to computing the area of the surface, the RT formula can be exactly reproduced in these models. Following the same idea, one can construct a PEE tensor network for subregions in AdS space and thereby build a toy model for holography restricted to a subregion. The factorized PEE tensor network provides the simplest setting in which the two conjectured formulations of holography for gravitational subregions, namely the surface/state correspondence \cite{Miyaji:2015yva} and the generalized entanglement wedge conjecture \cite{Bousso:2022hlz}, can be realized. These constructions may prove crucial for investigating quantum aspects of gravity when a standard holographic duality with an asymptotic boundary (and a dual QFT) is absent.

\begin{figure}[ht]
	\centering
	\includegraphics[width=0.5\linewidth]{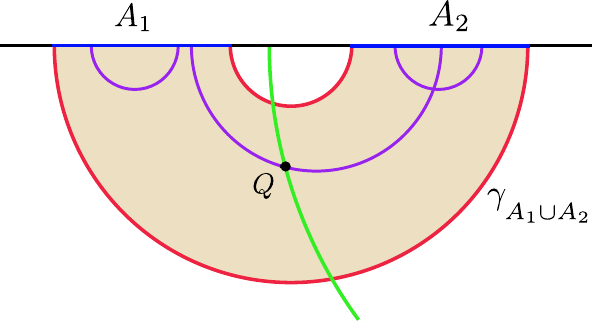}
	\caption{Failure of the factorized PEE tensor network for two boundary intervals $A_1$ and $A_2$. The green PEE thread passes through the interior of the entanglement wedge (shaded region) but does not anchor on $A_1\cup A_2$, causing the network to fail to reproduce the connected entanglement wedge.}
	\label{fig:disc}
\end{figure}

Nevertheless, the factorized PEE tensor network model is too simple to capture all properties of AdS/CFT. For example, our current approach fails to reproduce the entanglement wedge for multiple boundary intervals when the wedge is connected, even in a static slice of AdS$_3$. In the case of two intervals shown in Fig.\ref{fig:disc}, there exist PEE threads (see the green line) that pass through the interior of the entanglement wedge but do not anchor on the boundary intervals $A_1\cup A_2$. This issue might be resolved by upgrading the tensor network to the random tensor network \cite{Hayden:2016cfa} defined on the network of the PEE threads \cite{Wen:2025gui}.

It is also important and interesting to generalize the PEE tensor network from the vacuum AdS geometry to non-AdS spacetimes, as well as to configurations involving matter fields and black holes. We plan to introduce additional endpoints for the PEE threads in the bulk geometry and investigate whether a configuration of PEE threads can still form a network that tessellates the background geometry. These newly introduced endpoints would correspond to matter fields or black holes in the bulk. As a first step in this generalization, we will present a detailed discussion of the PEE configuration for the BTZ black hole in a separate paper \cite{upcoming}. The new endpoints of PEE threads are also responsible for the bulk quantum correction to the holographic entanglement entropy \cite{Faulkner:2013ana,Engelhardt:2014gca} and for the emergence of entanglement islands \cite{Penington:2019npb,Almheiri:2019psf,Almheiri:2020cfm}. We leave these topics for future investigations.

\acknowledgments

The authors are supported by the NSFC Grant No. 12447108 and the Shing-Tung Yau Center of Southeast University. We thank Yiwei Zhong and Yizhou Lu for helpful discussions and related collaborations.


	\bibliographystyle{JHEP}	\bibliography{Threads}
\end{document}